\documentclass[fleqn,usenatbib]{mnras}
\usepackage{newtxtext,newtxmath}
\usepackage[T1]{fontenc}
\DeclareRobustCommand{\VAN}[3]{#2}
\let\VANthebibliography\thebibliography
\def\thebibliography{\DeclareRobustCommand{\VAN}[3]{##3}\VANthebibliography}

\usepackage{graphicx}	% Including figure files
\usepackage{amsmath}	% Advanced maths commands
\usepackage{amsfonts}
\usepackage{float}
\usepackage{subcaption} % Sub-figures and Sub-tables
\usepackage{siunitx}    % Some special units
\usepackage{wrapfig}
\usepackage{adjustbox}
\usepackage{multirow}
\title[Lensing by 4D  EGB Bardeen black holes ]{Strong gravitational lensing by Bardeen black holes in 4D  EGB gravity: constraints from supermassive black holes}

\author[Shafqat Ul Islam, S.G. Ghosh and Sunil D. Maharaj]{
Shafqat Ul Islam $^{1}$\thanks{E-mail: shafphy@gmail.com}
, Sushant~G.~Ghosh$^{1,2}$ and Sunil D. Maharaj $^{2}$
\\
$^{1}$Centre for Theoretical Physics, Jamia Millia Islamia, New Delhi 110025, India\\
$^{2}$Astrophysics and Cosmology Research Unit, School of Mathematics, Statistics and Computer Science, University of KwaZulu-Natal, Private Bag 54001,\\ Durban 4000, South Africa
}

\begin{document}
\label{firstpage}
\pagerange{\pageref{firstpage}--\pageref{lastpage}}
\maketitle

% Abstract of the paper
\begin{abstract}
 Observation indicates that many nearby galaxies host supermassive central black holes. Modelling Bardeen models in four-dimensional Einstein-Gauss-Bonnet (4D EGB) gravity, with additional parameters $\tilde{\alpha}$ and charge $q$, as central black holes in various galaxies, we investigate gravitational lensing properties in strong deflection limits. Interestingly, the spherical photon orbit radius $x_m$, the critical impact parameter $u_m$,  the lensing coefficient $\bar{b}$, the deflection angle $\alpha_D(\theta)$, angular position $\theta_{\infty}$ are decreasing with  $q$ and  $\alpha$ whereas the other lensing coefficient $\bar{a}$ and  angular separation $s$ have opposite behaviour.  Taking the supermassive black holes Sgr A* and M87* as the lens, we also compare observable signatures of 4D EGB Bardeen black holes with those of the Schwarzschild black holes.  The angular position  $\theta_\infty$ for Sgr A* $\in$ (23.1853, \; 25.56427) $\mu$as, whereas for M87* it is $\in$ ( 17.941,\; 19.7819) $\mu$as. Further, the angular separation $s$, which is an increasing function of $\tilde{\alpha}$  and $q$ for Sgr A* and M87* differs significantly, respectively, in (0.031997,0.14895) $\mu$as and (0.0247, 0.1152) $\mu$as. The deviations of the lensing observables $\Delta \theta_{\infty}$ and $\Delta s$ for 4D EGB Bardeen black hole ($\tilde{\alpha}=0.9,~q=0.09$) from the Schwarzschild black hole, respectively,  can reach up to  $2.3789~\mu$as and $0.11695~\mu$as for Sgr A* , $1.84084~\mu$as and  $0.0905~\mu$as for M87*.  On the other hand, the relative magnification $\in$  (4.65751,\; 6.82173).  Considering twenty-two massive central black holes as  lens, we also estimate the time delay $\Delta T^s_{2,1}$  between the first and second relativistic image to find that, e.g.,  the time delay for Sgr A* and M87*,  respectively,  can reach $\sim9.86088$~min and $\sim16023.93$~min.  The latter is enough for any astronomical observation like Event Horizon Telescope. Further, we show that the shadow-size measurements place significant constraints on deviation parameters.   This combination of gravitational lensing and EHT results may complement comprehensive restrictions on modifications of the general relativity.  
\end{abstract}

% Select between one and six entries from the list of approved keywords.
% Don't make up new ones.

%%%%%%%%%%%%%%%%%%%%%%%%%%%%%%%%%%%%%%%%%%%%%%%%%%

%%%%%%%%%%%%%%%%% BODY OF PAPER %%%%%%%%%%%%%%%%%%

\section{Introduction}\label{Intro}
The singularities in Einstein's theory of general relativity (GR) appear to be a property inherent in most physically relevant exact solutions \citep{Hawking:1973}. Penrose (\citeyear{Penrose:1965}), assuming the weak energy condition and global hyperbolicity, proposed that formation of singularities in spacetime is inevitable. The cosmic censorship conjecture \citep{Penrose:1969pc} states that event horizons must surround these singularities, with no causal connection between the interior of a black hole and the exterior fields. However, it is wide belief that these singularities do not exist in Nature but are creations or artefacts of classical general relativity. By its very definition, the presence of singularity means spacetime fails to exist, signalling a breakdown of physics laws. Thus, other objects must substitute singularities in a more unified theory. The extreme condition  that may exist at the singularity imply that one should rely on quantum gravity, which are expanded to resolve these singularities \citep{Wheeler:1964,Tomozawa:2011gp,Cognola:2013fva,Zhang:2019dgi}. While we do not yet have a well-defined quantum gravity, to understand the inside of the black hole and resolve it separately; hence we must turn our attention to regular models, which are motivated by quantum arguments \citep{Ansoldi:2008jw}.   Bardeen (\citeyear{Bardeen:1968}), the first to introduce the regular black hole model, which has an event horizon with no singularities, has a de-Sitter core. Hence, the Bardeen model is the most significant regular black hole, triggering a flare-up of regular black hole research activities, which include investigating thermodynamical properties \citep{12}, geodesics equations \citep{Stuchlik:2014qja}, quasinormal modes \citep{Fernando:2012yw}, Hawking evaporation \citep{Mehdipour:2016rtd} and black hole remnant \citep{Mehdipour:2016rtd}. Lately, Bardeen's solution has been extended to higher-dimensional spacetime \citep{Ali:2018boy,Kumar:2018vsm}, and to its rotating counterpart \citep{Bambi:2013ufa,Ghosh:2015pra,Ali:2019myr}. Thus, there has been immense advancement in the analysis and application of Bardeen's black holes in GR.  It motivated Bardeen's model and it's applications in Einstein-Gauss-Bonnet (EGB) theories of gravity \citep{Schee:2015nua,Kumar:2018vsm,Singh:2019wpu,Kumar:2020uyz}.

EGB theory  is one of the  simplest natural extensions of GR to higher dimensions $D \ge 5$. The first spherically symmetric static black hole solution in EGB gravity was obtained by  \cite{David1985}  and analysed latter in cascade of works \citep{Wiltshire:1988uq,ms,Kanti:1996gs,Sahabandu:2005ma,Ghosh:2008jca,Ghosh:2011ad,Ghosh:2014pga,Antoniou:2017acq,Antoniou:2017hxj,Bakopoulos:2018nui}. The GB correction to the Einstein Hilbert action in $D=4$ is a total derivative and gives a non trivial contribution to the gravitational dynamics.  The 4D EGB gravity in \citep{Glavan:2019inb}  is on equal footing with the GR and has received an substantial attention which include  Vaidya-like radiating black holes \citep{Ghosh:2020vpc}, charged spherically symmetric black holes \citep{Fernandes:2020rpa}, rotating black holes  \citep{Kumar:2020owy,Wei:2020ght},  gravitational lensing \citep{Islam:2020xmy,Kumar:2020sag,Jin:2020emq}, thermodynamical properties of anti-de Sitter black holes \citep{Konoplya:2020bxa} and other contributions  \citep{Ghosh:2020syx,Singh:2020mty,Kumar:2020xvu}. 

Gravitational lensing, a vital application of GR, is a general term used to account for all effects of the gravitational field on the propagation of electromagnetic radiation. Light propagates along a straight path unless an object with a gravitational field intervenes. Gravitational lensing is an  essential astrophysical tool to extract information about distant stars, highly red-shifted galaxies, quasars, supermassive black holes, and exoplanets that are otherwise too dim to be observed. It has been used to determine the Hubble constant \citep{Dyer:1980}, probe the structure of galaxies \citep{Astron:1989}, dark matter and dark energy in galactic halos \citep{Chang:1979}, measure the density of cosmic string \citep{Gott:1986}.  

The weak-field limit, based on the assumption that the deflection angle is small, has been extensively studied \citep{Schneider:1992,Blandford:1992,Zakharov:1997}, and it has successfully explained the experimental tests done on GR. However, in the vicinity of compact objects like a black hole where the lensing will have a rich structure, GR is yet to be tested. Darwin (\citeyear{Darwin:1959}) and Schneider (\citeyear{Schneider:1992})  were the first to notice that light rays would make one or more loops around the black hole, resulting in an infinite sequence of exotic images. Some years later, in what we could consider the beginning of black hole imaging, Walsh (\citeyear{Walsh:1965}) discovered the first example of gravitational lensing in which they reported multiple images of a binary Quasar. Following the remarkable discovery of Quasars,  gravitational lensing in strong deflection limit (SDL) was extensively studied \citep{Bhadra:2003zs}. The gravitational lensing in SDL was though resurrected by  Virbhadra and Ellis (\citeyear{Virbhadra:1999nm}), who found the exact lens equation with a large deflection angle for the galactic supermassive black hole in an asymptotically flat background. Frittelli  (\citeyear{Frittelli:1999yf}) used an approach to construct the lens equation and time of arrival without reference to a background metric for a Schwarzschild black hole.  \cite{Bozza:2001xd}, developed an analytical method to investigate Schwarzschild black hole lensing in  SDL and found a logarithmic divergence of the deflection angle. They have extended this technique to Reissner black holes  \citep{Eiroa:2002mk}, and more so to an arbitrary static spherically symmetric metric  \citep{Bozza:2002zj}. The studies in gravitational lensing in SDL achieved a real boost when the first image of black hole M87* \citep{EventHorizonTelescope:2019dse,EventHorizonTelescope:2019uob,EventHorizonTelescope:2019jan,EventHorizonTelescope:2019ths,EventHorizonTelescope:2019pgp,EventHorizonTelescope:2019ggy} was captured by the Event Horizon Telescope (EHT). Black holes have turned into a physical reality in 2019 with the release of the first horizon-scale image of the supermassive black hole M87* by the Event Horizon Telescope (EHT) collaboration \citep{EventHorizonTelescope:2019dse,EventHorizonTelescope:2019pgp,EventHorizonTelescope:2019ggy}.  Using a distance of $d=16.8$ Mpc and estimated mass of M87*  $M=(6.5 \pm 0.7) \times 10^9 M_\odot$ \citep{EventHorizonTelescope:2019dse,EventHorizonTelescope:2019pgp,EventHorizonTelescope:2019ggy}, the EHT collaboration declared compact emission region size with angular diameter $\theta_d=42\pm 3\, \mu $as  with the central flux depression with a factor of $\gtrsim 10$, which is the black hole shadow. 
It took a further lift with the recent EHT announcement of the Sgr A* black hole shadow results, showing shadow angular diameter $\theta_{sh}= 48.7 \pm 7\,\mu$as with enveloping bright and thick emission ring of diameter $\theta_d=51.8\pm 2.3\mu$as \citep{EventHorizonTelescope:2022xnr,EventHorizonTelescope:2022urf,EventHorizonTelescope:2022xqj}. The EHT assumed Sgr A* black hole of mass $M = 4.0^{+1.1}_{-0.6} \times 10^6 M_\odot $  and distance $D=8$kpc from earth, the EHT show that the Sgr A* black hole shadow are persistent with the expected appearance of a Kerr black holes \citep{EventHorizonTelescope:2022exc,EventHorizonTelescope:2022xqj}. When compared with the EHT results for M87*, it exhibits consistency with the predictions of GR \citep{EventHorizonTelescope:2022xnr}. It also opens the gateway to investigate the region near the black hole and to test the validity of alternate theories of gravity \citep{Kumar:2018ple,Kumar:2020yem,Kumar:2020ltt,Afrin:2021imp,Afrin:2021wlj}.   This has motivated us to investigate the gravitational lensing by Bardeen black holes in 4D EGB gravity $\sim$ 4D EGB Bardeen black holes. We calculate the observables in terms of lensing coefficients and determine the effect of charge and coupling constant on the observables. 

The rest of the paper is organized as follows: in Sec.~\ref{sect2},  we introduce the 4D EGB Bardeen black holes which are  necessary to efficiently calculate  properties of photon in SDL, including a discussion on the parameter space for black holes and horizon structures. A discussion on the gravitational lensing of light in  SDL for 4D EGB Bardeen black holes is the subject of  Sec.~\ref{sect3}. Sec.~\ref{sect5} is devoted to the evaluation  of lensing observables by 4D EGB Bardeen black holes, including the image positions, separation and magnifications  by  supermassive black holes Sgr A* and M87*. By taking the supermassive black holes as the lens, we numerically estimate time delays of the images in Sec.~\ref{td}. In Sec.~\ref{sec6} we estimate and compare the lensing observables with the GR counterparts. Finally, we summarize our main findings in  Sec.~\ref{sect6}. 
 
Throughout this paper, unless otherwise stated, we adopt natural units ($8\pi G = c = 1$)
\begin{figure}
\begin{center}	
\includegraphics[scale=0.77]{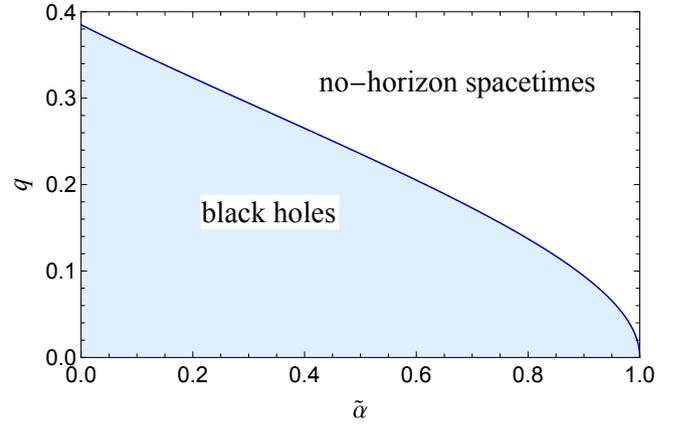}
\caption{The parameter space for 4D EGB Bardeen black holes. The dark-blue line corresponds to values of parameters for extremal 4D EGB Bardeen black holes. } \label{plot1} 
\end{center}  
\end{figure}
\section{4D EGB Bardeen Black Holes}\label{sect2}
Lovelock (\citeyear{Lovelock:1971yv}) demonstrated that Einstein gravity could be extended by series of higher order curvature terms such that the equation of motion still remains second order. The simplest such extension is   EGB gravity when coupled with the non-linear electrodynamics. The action reads \citep{Cai:2001dz,  Glavan:2019inb,Ghosh:2020tgy}  
\begin{equation} \label{action}
\mathcal{I}_{G}=\frac{1}{2}\int_{\mathcal{M}}d^{D}x\sqrt{-g}\left[  R +\alpha\mathcal{L}_{GB} +\mathcal{L}(F) \right],
\end{equation}
where $\alpha$ is the coupling constant, $g$ is the  determinant of metric tensor $g_{\mu\nu}$ and $R$ is the Ricci scalar. The Gauss-Bonnet lagrangian is a combination of Ricci tensor $R_{\mu \nu}$ and Riemann tensor  $R_{\mu \nu \rho \sigma}$ given by
\begin{eqnarray}
\mathcal{L}_{GB}&=& R^2 -4 R_{\mu \nu}R^{\mu \nu}+R_{\mu \nu \rho \sigma}R^{\mu \nu \rho \sigma},
\end{eqnarray}
and $\mathcal{L(F)}$ is an arbitrary function of the invariant $\mathcal{F}=F_{\mu\nu}F_{\mu\nu}$, where $F_{\mu\nu}=\partial_{\mu}A_{\nu}-\partial_{\nu}A_{\mu}$ is the electromagnetic field tensor for the gauge potential $A_{\mu}$.  When $D=4$, the $\mathcal{L}_{GB}$ becomes a topological invariant and hence does not contribute to the dynamics \citep{Glavan:2019inb}. The variation of the action (\ref{action}) yields field equations in the form
\begin{equation}\label{fe} 
G_{\mu\nu} +{\alpha}H_{\mu\nu} = T_{\mu\nu}
\equiv 2\Bigr[\mathcal{L_F} \mathcal{F}_{\mu\sigma}\mathcal{F}{^\sigma}_{\nu}-\frac{1}{4}g_{\mu\nu}\mathcal{L(F)}\Bigl],
\end{equation}
where the Einstein tensor $G_{\mu\nu}$ and Lanczos tensors $H_{\mu\nu}$ \citep{Lanczos:1938sf}, respectively, are given by 
\begin{eqnarray}
G_{\mu\nu}&=&R_{\mu\nu}-\frac{1}{2}Rg_{\mu\nu},
\\
H_{\mu\nu}&=&2(RR_{\mu\nu}-2R_{\mu\sigma}R_{\ \nu}^{\sigma}-2R_{\mu\sigma\nu\rho}R^{\sigma\rho}
\nonumber\\&-&R_{\mu\sigma\rho\beta}R_{\nu}^{\sigma\rho\beta}) - \frac{1}{2}g_{\mu\nu}\mathcal{L}_{GB}.
\end{eqnarray}
\begin{figure*}
\begin{center}	
\begin{tabular}{p{9cm} p{9cm}}
\includegraphics[scale=0.62]{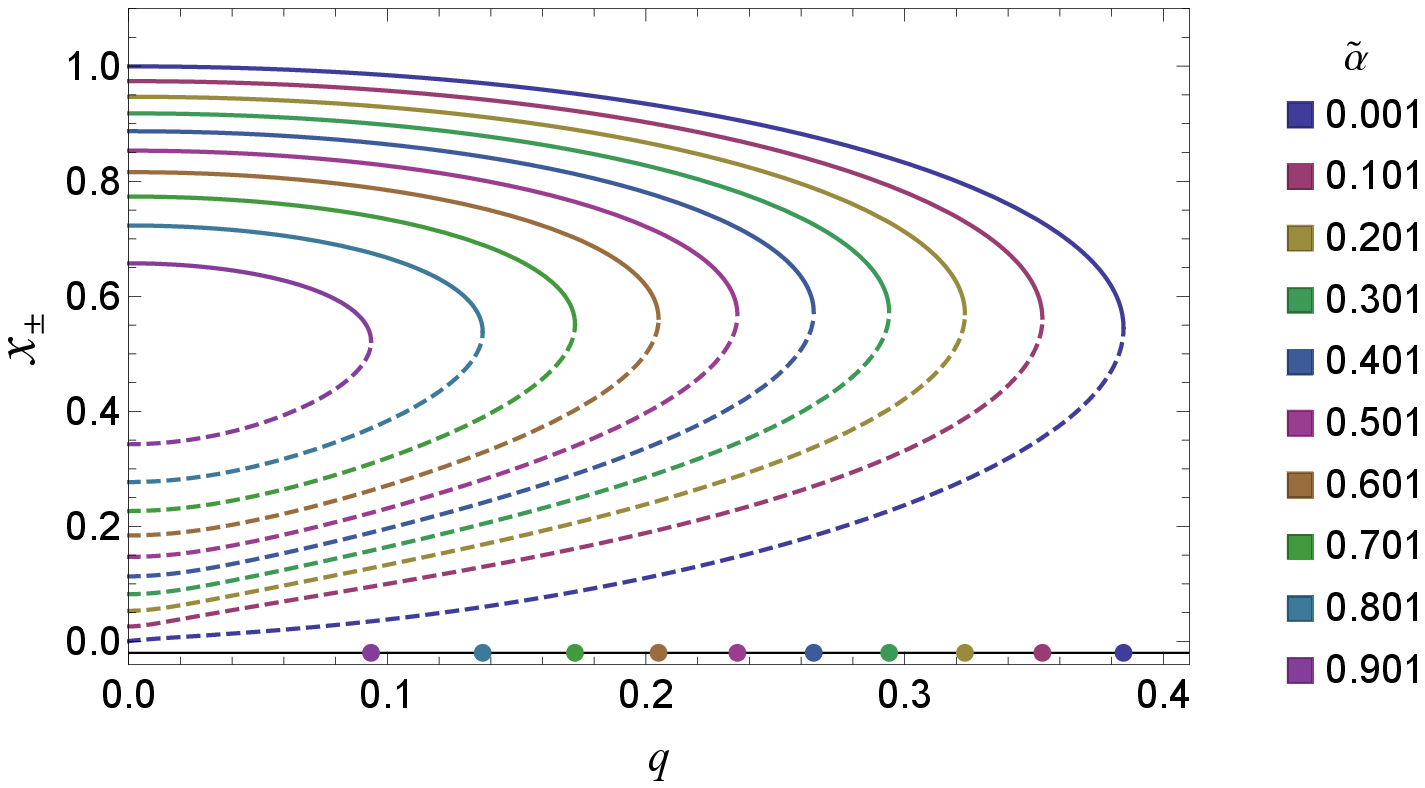}&
\includegraphics[scale=0.62]{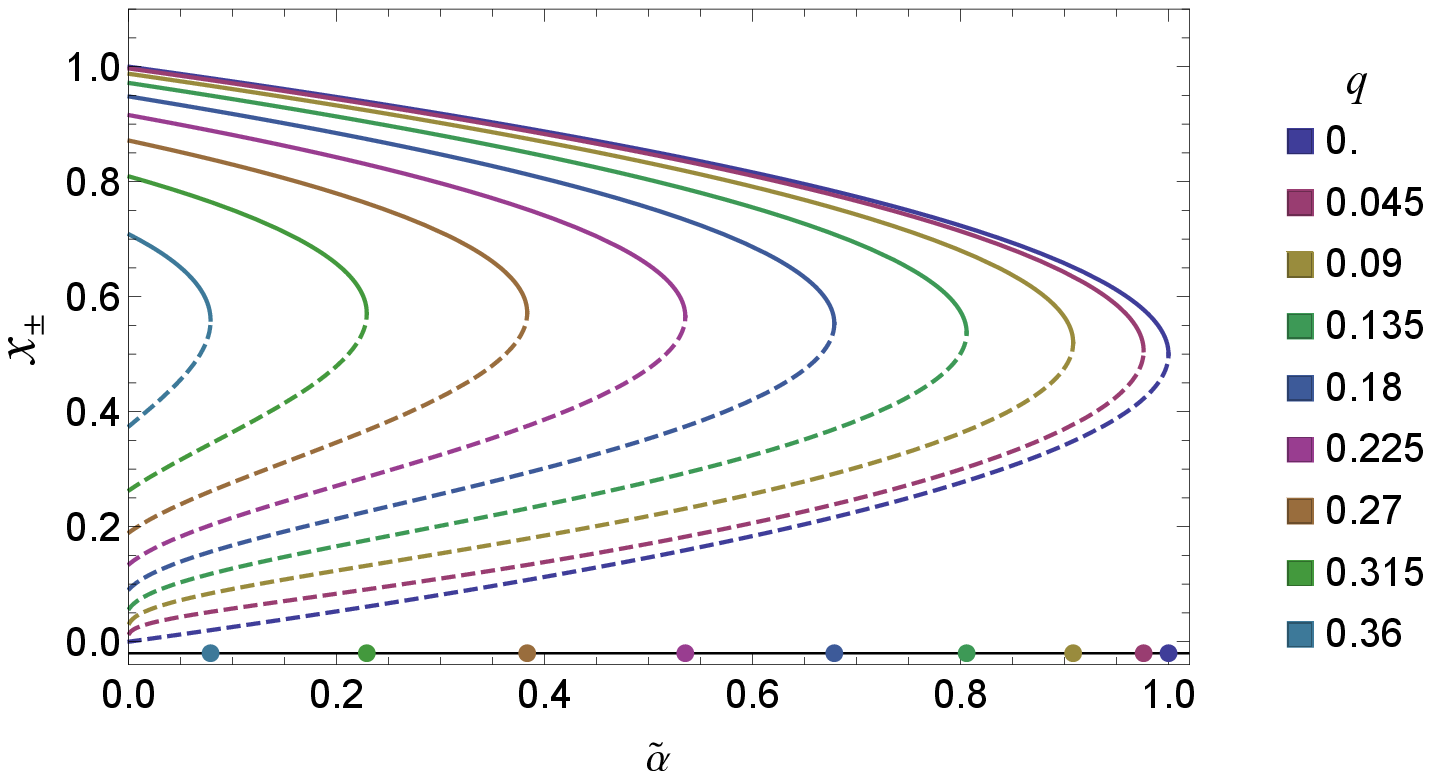}\\
\end{tabular}
\caption{The behaviour of  event horizon radii (solid lines), Cauchy horizon radii (dashed lines) as a function of   $q$ for different values of  $\tilde{\alpha}$ (\textit{left}) and as a function of  $\tilde{\alpha}$ for different values of $q$ (\textit{right}). The bullet points on the $x$-axis correspond to  extremal values. Our results in the limit $\tilde{\alpha}\to 0$ encompass those of Bardeen black holes, those of Schwarzschild black hole when $\tilde{\alpha}\to 0,~q=0$, and if $q=0$, $\tilde{\alpha}\ne 0$, we obtain  results of  the 4D EGB black holes.}	\label{plot2}
\end{center}  
\end{figure*}
\begin{figure*}
	\begin{centering}
		\begin{tabular}{p{10cm} p{10cm}}
		\includegraphics[scale=0.8]{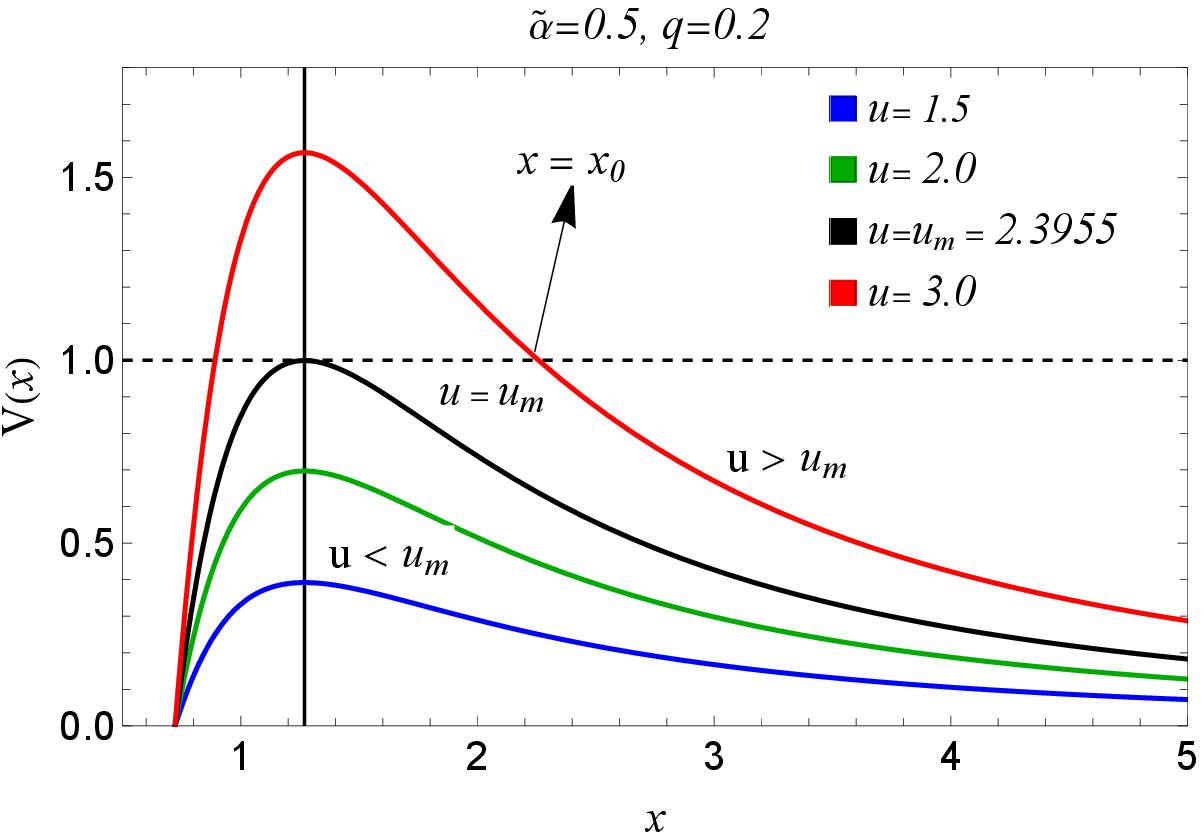}&
		\includegraphics[scale=0.8]{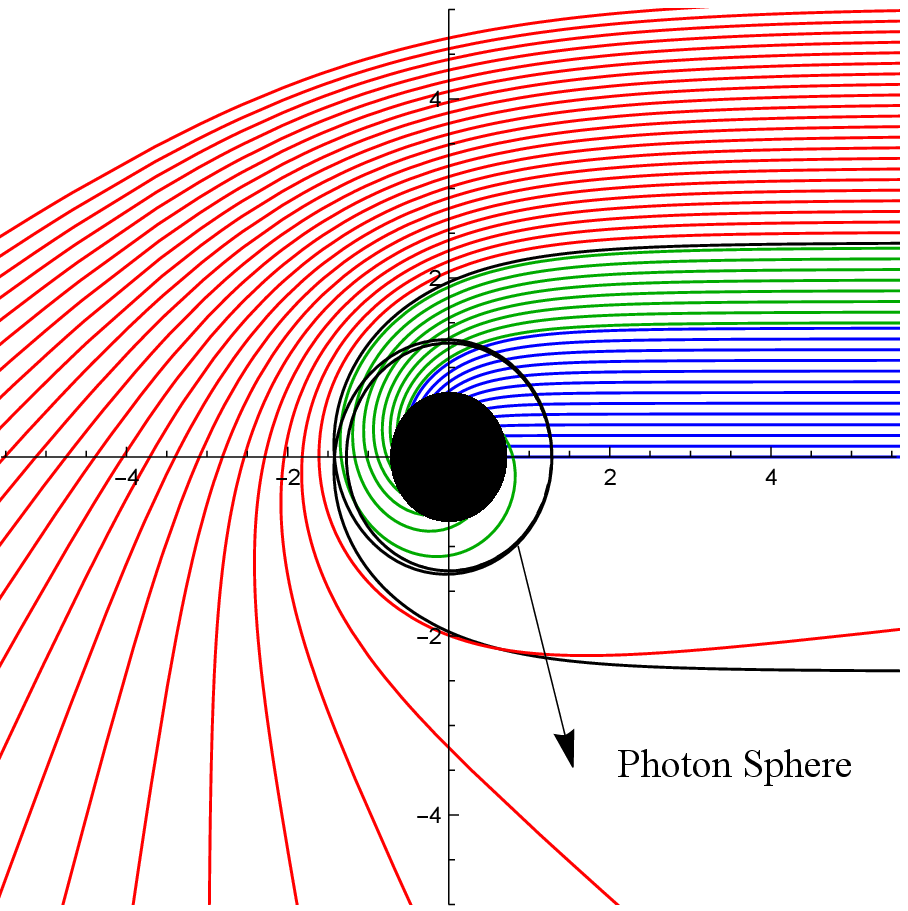}
			\end{tabular}
	\end{centering}
	\caption{ Effective potential $V$ as a function of $x$ for $q=0.2$ and $\tilde{\alpha}=0.5$. The trajectories with $u>u_m$, $u=u_m$ and $u>u_m$ are represented by red, black, green (and blue) respectively (\textit{left}). The trajectory of the light ray for  $ \tilde{\alpha}=0.5 $, $q=0.2$, in  polar coordinates ($r,\phi$). The black line corresponds to the value of  $u$ close to  $u_m$. The black hole is shown as a solid disk and the photon sphere  as a dashed black circle (\textit{right}).} \label{plot5}		
\end{figure*}  
We wish to obtain  Bardeen-like black holes from Eq.~(\ref{fe}) for which the Lagrangian density reads \citep{Ali:2018boy,Kumar:2018vsm}
\begin{equation}
\mathcal{L(F)}=\frac{(D-1)(D-2)\mu^{D-3}}{4g^{D-1}}\left(\frac{\sqrt{2g^2 \mathcal{F}}}{1+\sqrt{2g^2 \mathcal{F}}} \right)^{\frac{2D-3}{D-2}},
\end{equation} 
with  $\mathcal{F}={g^{2(D-3)}}/{2r^{2(D-2)}}$.
Here we choose the general static, spherically symmetric metric anstaz in arbitary dimensions  \citep{Kumar:2018vsm,Ghosh:2020tgy} as
\begin{equation}\label{metric}
ds^2=-A(x)dt^2+\frac{dx^2}{A(x)}+ x^2 d\Omega_{D-2}^2,
\end{equation} 
where $d\Omega^2_{D-2}$ is the line element of a $(D-2)$-dimensional constant curvature space \citep{Myers:1986un}.  Using field equations (\ref{fe}) for the metric ansatz (\ref{metric}), rescaling $\alpha\to\alpha/(D-4)$  \citep{Glavan:2019inb}, and then taking the limit $D \to 4$,  we obtain the solution \citep{Kumar:2020uyz}
\begin{equation} \label{fr}
A_{\pm}(x)=1 + \frac{2 x^{2}}{\tilde{\alpha}}\left(1 \pm \sqrt{1+\frac{\tilde{\alpha}}{( x^2+q^2)^{3/2}}}\right),
\end{equation}
by appropriately relating the constant of integration with the black hole mass $M$. We have also used  $q=g/2M$, $\tilde{\alpha}=\alpha/M^2$ and measured all lengths in units of  radius $2M$.  Eq.~(\ref{fr}) represents the Bardeen-like black holes in 4D EGB gravity and henceforth, referred to as 4D EGB Bardeen black holes.

In the limit $\tilde{\alpha}\to 0$, the $-$ve branch corresponds to  Bardeen black holes \citep{Bardeen:1968}, whereas $+$ve branch leads to an unphysical solution, and  hence we shall restrict ourselves to the $-$ve branch. Further, in the limit $x\to\infty$, the $-$ve branch  is asymptotically flat. Interestingly, the gravity with quantum corrections  \citep{Cognola:2013fva} and the semi-classical Einstein equations with conformal anomaly \citep{Cai:2009ua} also admit the solution Eq.~(\ref{fr}). The 4D EGB Bardeen black holes (\ref{fr}) encompass the Bardeen black holes when $\tilde{\alpha}\to 0$ and 4D EGB black holes \citep{Glavan:2019inb} when $q=0$. When $\tilde{\alpha}\to 0$ and $q=0$, it resembles the Schwarzschild black hole. 

The 4D EGB Bardeen black holes (\ref{fr}) are characterised by the parameters $(\tilde{\alpha}, q)$  and its horizons are the zeroes of $g^{xx}=A(x)=0$, which for $q=0$ admit
\begin{eqnarray}
x_{\pm}=\frac{1}{2}\left(1 \pm\sqrt{1-\tilde{\alpha}} \right)
\end{eqnarray}
In general, $A(x) = 0$ admits two roots $x_{\pm}$, when parameters ($\tilde{\alpha}$, $q$) are in the blue region of parameter space (cf. Fig.~\ref{plot1}), which correspond to the inner Cauchy horizon ($x_{-}$) and the outer event horizon ($x_{+}>x_{-}$) (cf. Fig.~\ref{plot2}). When the parameters ($\tilde{\alpha}$, $q$) are on the dark-blue line in the Fig.~\ref{plot1}, we have  extremal 4D EGB Bardeen black holes with $x_{e}=x_{+}=x_{-}$.  In Fig.~\ref{plot2}, we have shown the dependence of the event horizon and the Cauchy horizon on $\tilde{\alpha}$ and $q$. It is evident from Fig.~\ref{plot2}  that there are  extremal values of $q_e$ ($\tilde{\alpha}_e$) represented by  bullet points on the  $x$-axis  for given values of $\tilde{\alpha}$ ($q$) such that for $q < q_e $ ($\tilde{\alpha}<\tilde{\alpha}_e$) there exists two  horizons $(x_{\pm})$. 
\begin{figure*}
\begin{center}	
\begin{tabular}{p{9cm} p{9cm}}
\includegraphics[scale=0.62]{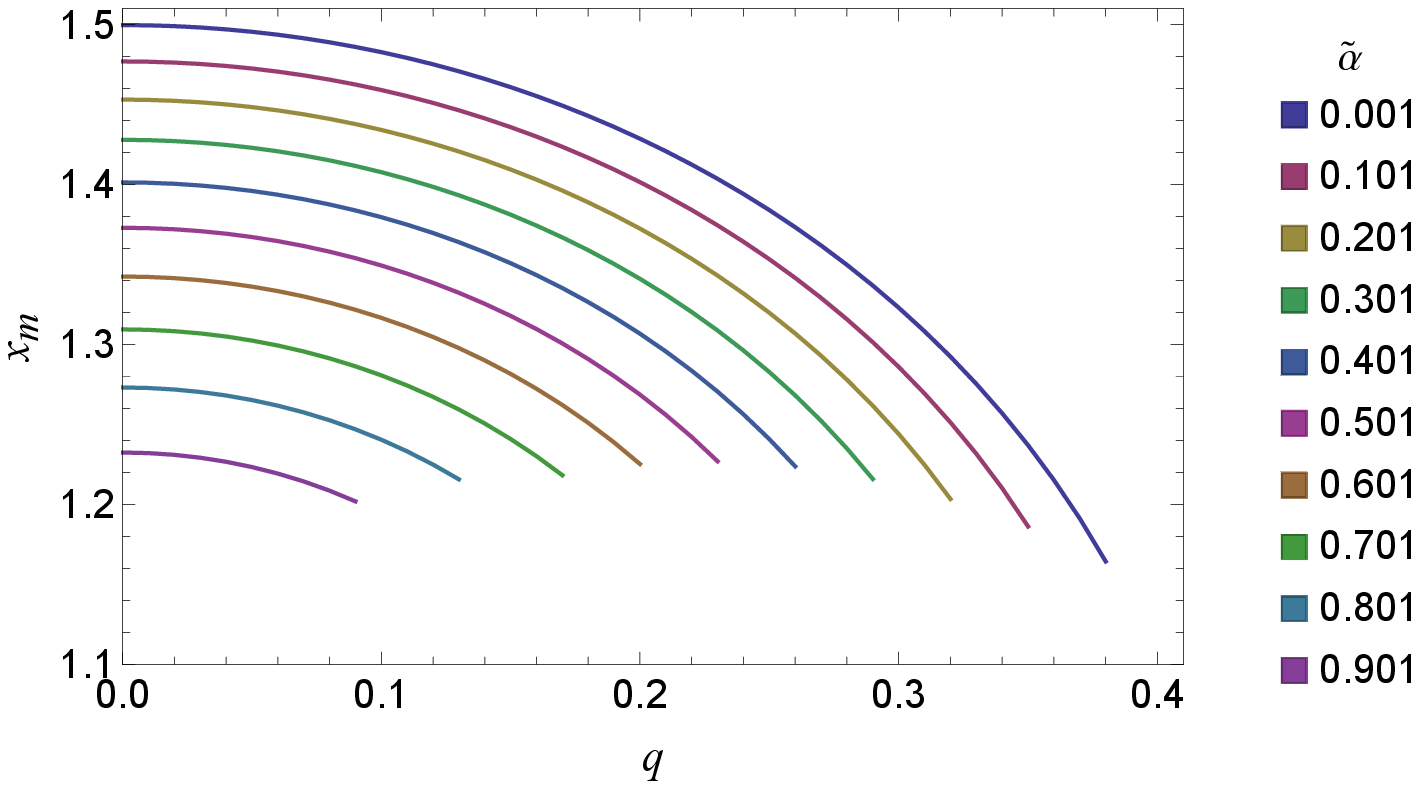}&
\includegraphics[scale=0.62]{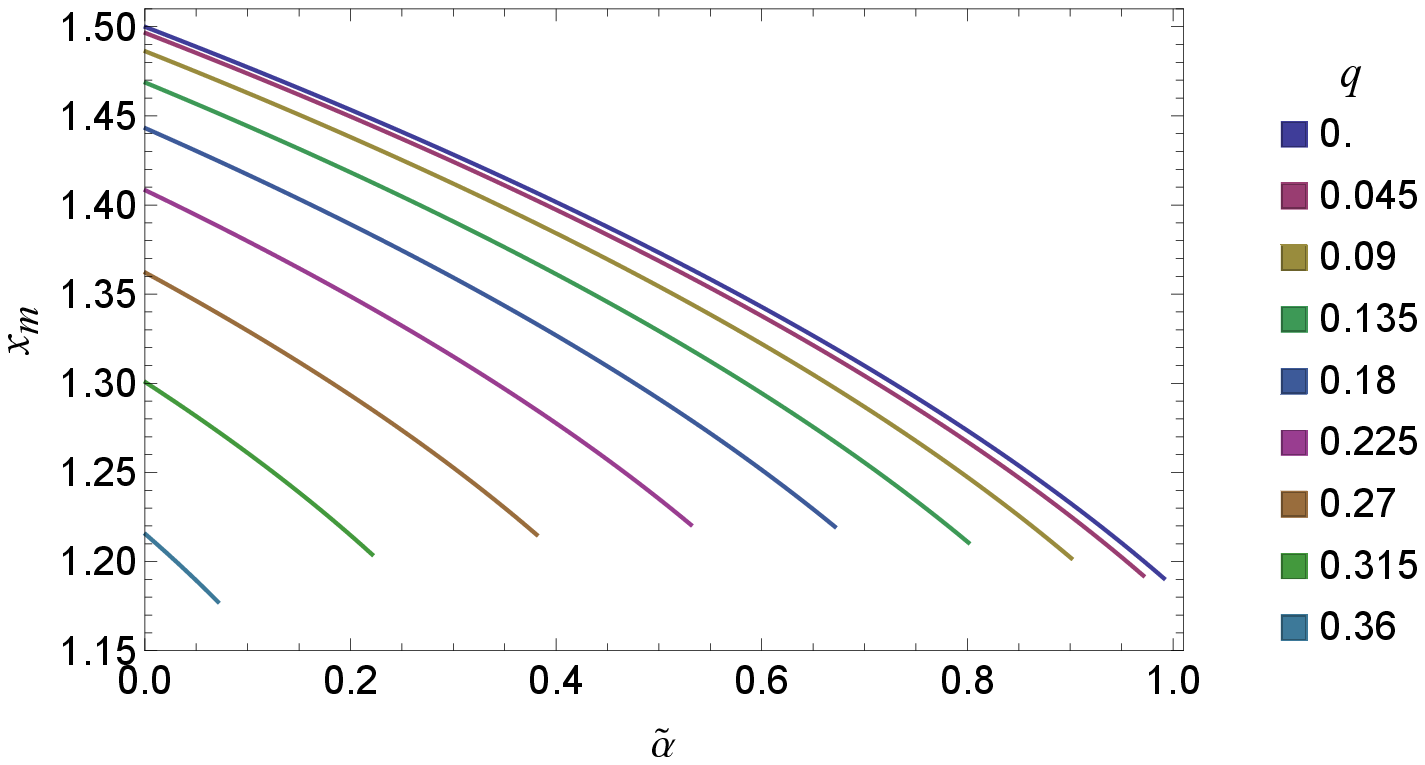}
\end{tabular}
\caption{The variation of unstable photon orbit radius as a function of   $q$ for different values of  $\tilde{\alpha}$ (\textit{left}) and as a function of  $\tilde{\alpha}$ for different values of $q$ (\textit{right}).  Our results in the limits $\tilde{\alpha}\to 0$ encompass those of Bardeen black holes, those of Schwarzschild black hole when $\tilde{\alpha}\to 0,~q=0$ and if $q=0$, $\tilde{\alpha}\ne 0$, we obtain  results of the 4D EGB black holes.}\label{plot3}
\end{center}  
\end{figure*}

Before we start our discussion on  gravitational lensing in SDL, we  clarify that the \textit{regularisation} proposed in \cite{Glavan:2019inb} and \cite{Cognola:2013fva}, is subject to dispute and many authors raised questions \citep{Gurses:2020ofy,Hennigar:2020lsl,Ai:2020peo,Shu:2020cjw,Mahapatra:2020rds,Arrechea:2020evj}. Many alternative \textit{regularisation} have also been suggested 
\citep{Lu:2020iav,Kobayashi:2020wqy,Hennigar:2020lsl,Fernandes:2020nbq,Casalino:2020kbt}. However, the spherically symmetric 4D black hole solution obtained  \citep{Cognola:2013fva,Glavan:2019inb,Ghosh:2020syx} still remains valid in these regularised theories \citep{Lu:2020iav,Hennigar:2020lsl,Fernandes:2020nbq,Casalino:2020kbt}. Hence these  \textit{regularisation} procedure lead to exactly the same black hole solutions \citep{Cognola:2013fva,Glavan:2019inb} at least for the case of 4D spherically symmetric spacetimes. We can confirm that our solution (\ref{fr}) can be obtained by the  \textit{regularisation} proposed by \cite{Hennigar:2020lsl}.
\begin{figure}
\begin{center}	
\includegraphics[scale=0.6]{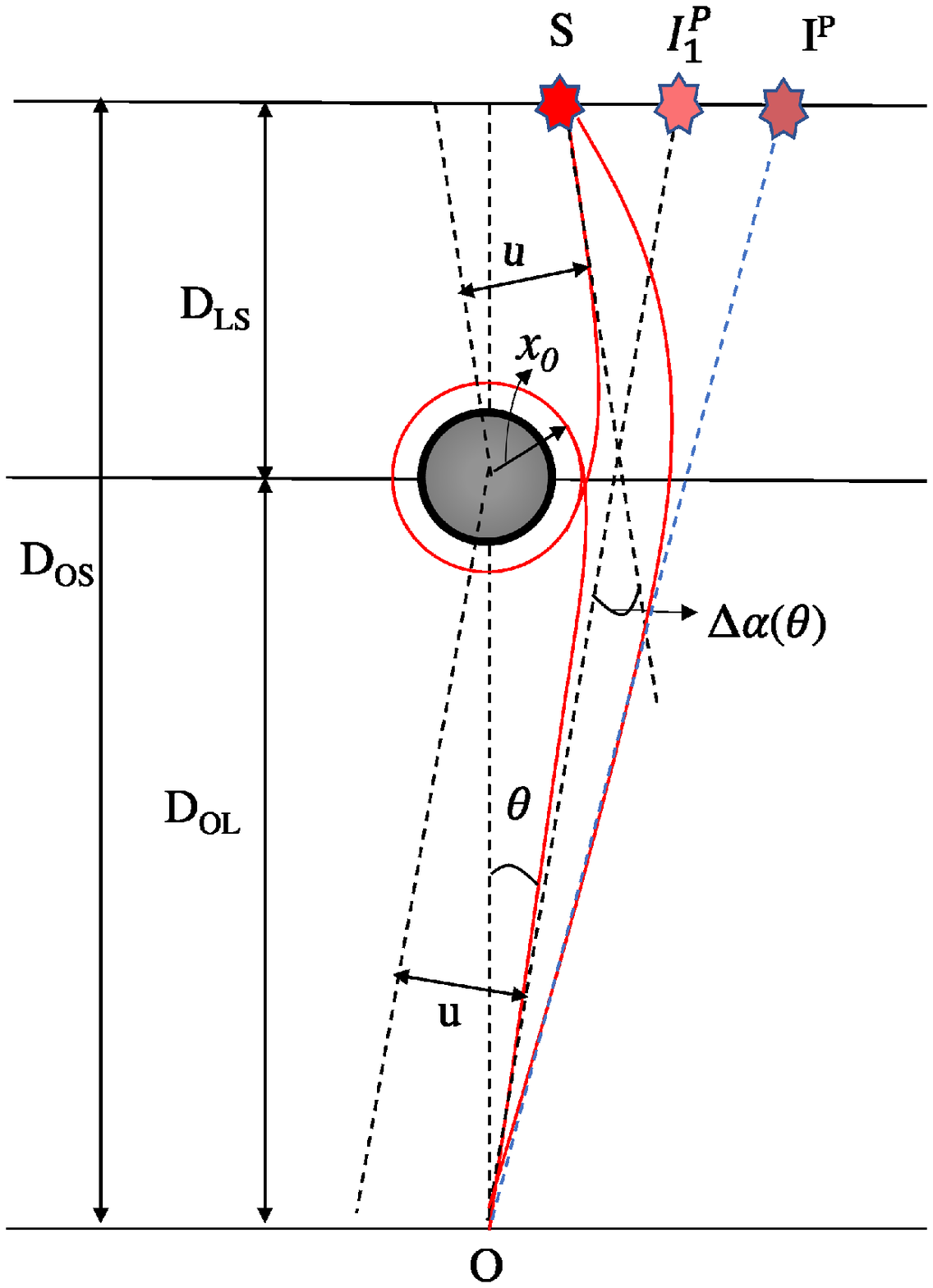}
\caption{Formation of primary images and relativistic images of source space(S) in case of gravitational lensing. Light rays are deviated by the black hole to be observed at an angular position $\theta$ by the observer space(O). } \label{plot5b} 
\end{center}  
\end{figure}
\section{Gravitational lensing in SDL}\label{sect3}
After the above discussion on the  spherically symmetric 4D EGB Bardeen black holes, we now  consider the gravitational  lensing  in SDL. The black hole significantly influences the motion of a photon in its neighbourhood (cf.~Fig.~\ref{plot5}).  The photon from a source approaches the black hole to a minimum distance $x_{0}$ and is deflected by its  gravitational field to be received by the observer at infinity (cf.~Fig.~\ref{plot5b}). The light ray trajectory is described by $k^{\mu}k_{\mu}=0$, where $k^{\mu}=\dot{x}^{\mu}$ is a wave number of light and the overdot denotes  differentiation with respect to the affine parameter along the trajectory \citep{Tsukamoto:2016jzh}. The trajectories  of the light rays are depicted in Fig.~\ref{plot5}. The energy $\mathcal{E}\equiv-g_{\mu\nu}t^{\mu}k^{\nu}=A(x)\dot{t}$
and angular momentum $\mathcal{L}\equiv g_{\mu\nu}\phi^{\mu}k^{\nu}=x^2\dot{\phi}$ are constant along the light trajectory. From $k^{\mu}k_{\mu}=0$, we obtain the trajectories of the light as  \citep{Chandrasekhar:1985kt,Islam:2020xmy,Kumar:2021cyl}
\begin{equation}\label{traj}
-A(x)\Dot{t}^2 + \frac{\Dot{x}^2}{A(x)} + C(x) \Dot{\phi}^2=0,    
\end{equation}
where $C(x)=x^2$. Using the definitions of energy $\mathcal{E}$, angular momentum $\mathcal{L}$ and introducing the impact parameter $u=\mathcal{L}/\mathcal{E}$,  Eq.~(\ref{traj}) can be rewritten as \citep{Chandrasekhar:1985kt,Eiroa:2010wm,Tsukamoto:2020bjm}
\begin{equation}
    \Dot{x}^2+V(x)=1,~~~~~~~~V(x)=u^2\frac{A(x)}{C(x)},
\end{equation}
where $V(x)$ is its effective potential (cf. Fig.~\ref{plot5} for its behaviour).  The  unstable spherical photon orbits radii  which correspond to the distance where deflection angle diverges, are timelike hypersurfaces obtained by simultaneously solving $\dot{x}=\ddot{x}=0$ for $V''(x)<0$ giving $C'(x)/C(x)=A'(x)/A(x)$ \citep{Bozza:2002zj,Tsukamoto:2016jzh,Eiroa:2010wm,Islam:2020xmy,Kumar:2021cyl},  whose radius leads to the unstable photon sphere radius by $x_m$ and it is depicted  for 4D EGB Bardeen black holes with varying $\tilde{\alpha}$ ($q$)  in Fig.~\ref{plot3} right (left). The photon orbit radius is a monotonically decreasing function of $\tilde{\alpha}$ and $q$, particularly, $x_m$ of Bardeen black holes as well as 4D EGB black holes, is always larger than the  4D EGB Bardeen black holes \citep{Eiroa:2010wm}. Moreover, the  radius $x_m$ peaks to a value of $1.5$  when $\tilde{\alpha}\to0$ and $q=0$ \citep{Bozza:2002zj,Eiroa:2002mk,Eiroa:2010wm}. Without loss of generality, we set $\mathcal{E}=1$ such that at the distance of minimum approach $x_0$, which is by definition the turning point of the photon trajectory i.e., $\Dot{x}=0$ \citep{Bozza:2002zj,Eiroa:2010wm,Tsukamoto:2016jzh,Islam:2020xmy}, the impact parameter reads 
\begin{equation}\label{angmom}
u^2 = \frac{x_0^2 \tilde{\alpha}}{2x_0^2(1-\sqrt{1 +  \tilde{\alpha}/(q^2 + x_0^2)^{3/2} } )+\tilde{\alpha}}.    
\end{equation}
\begin{figure*}
	\begin{centering}
		\begin{tabular}{p{9cm} p{9cm}}
		    \includegraphics[scale=0.62]{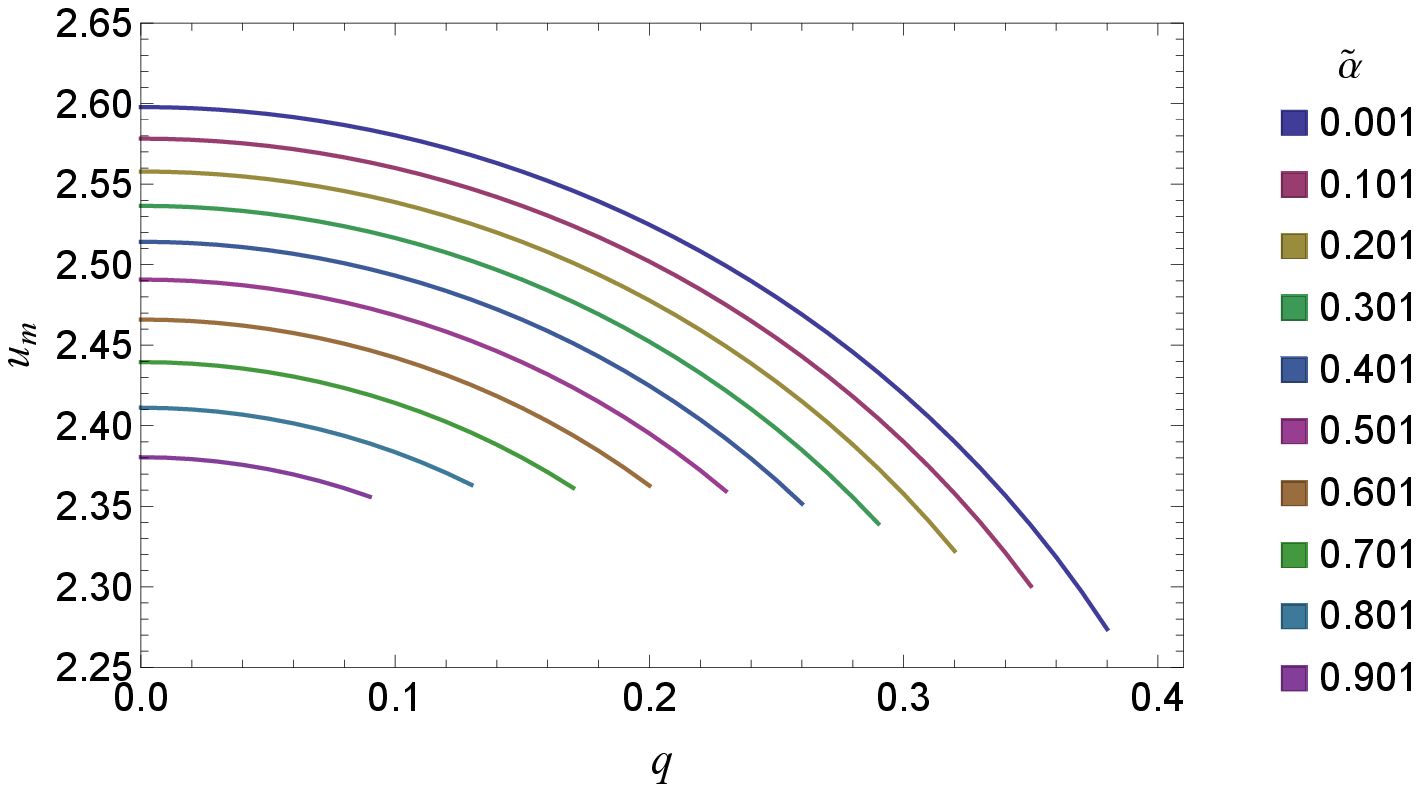}&
			\includegraphics[scale=0.62]{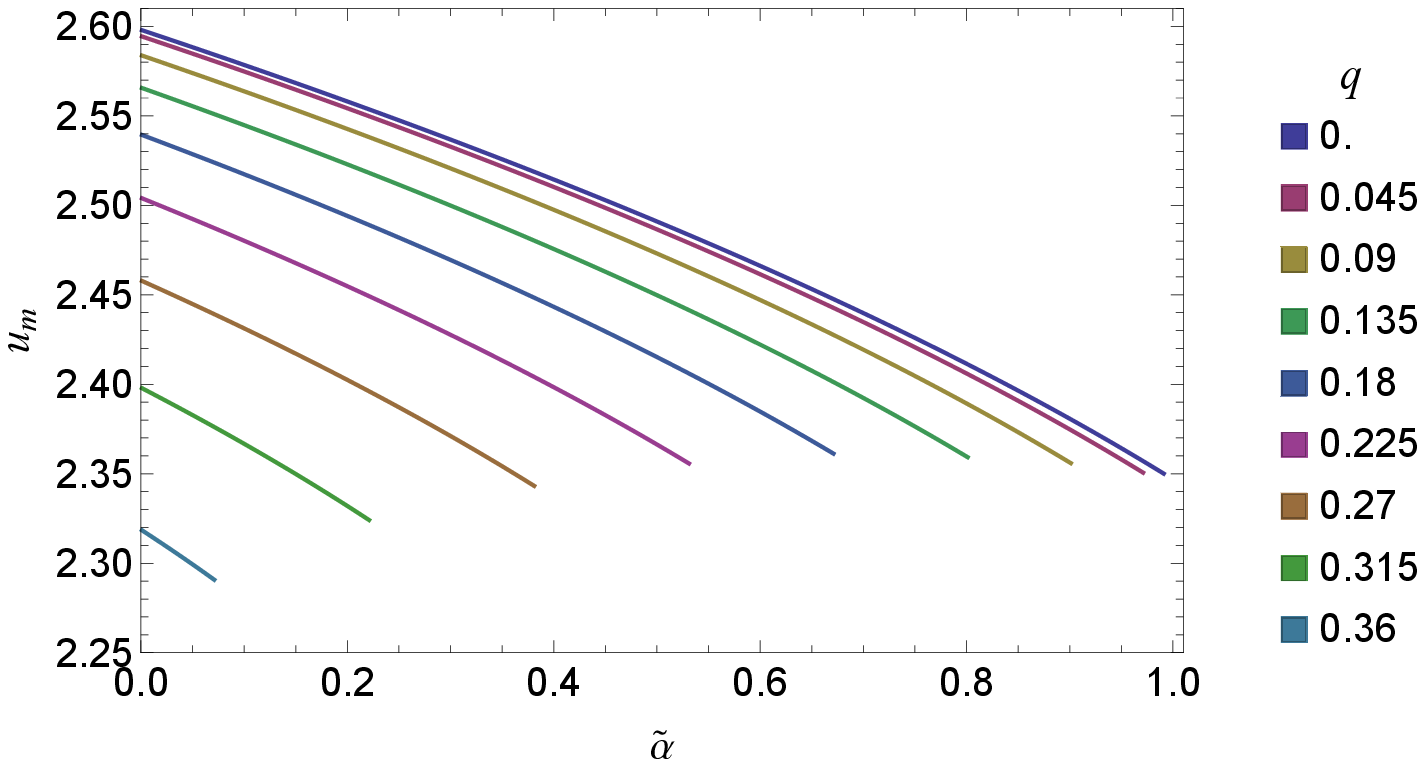}\\
			\includegraphics[scale=0.62]{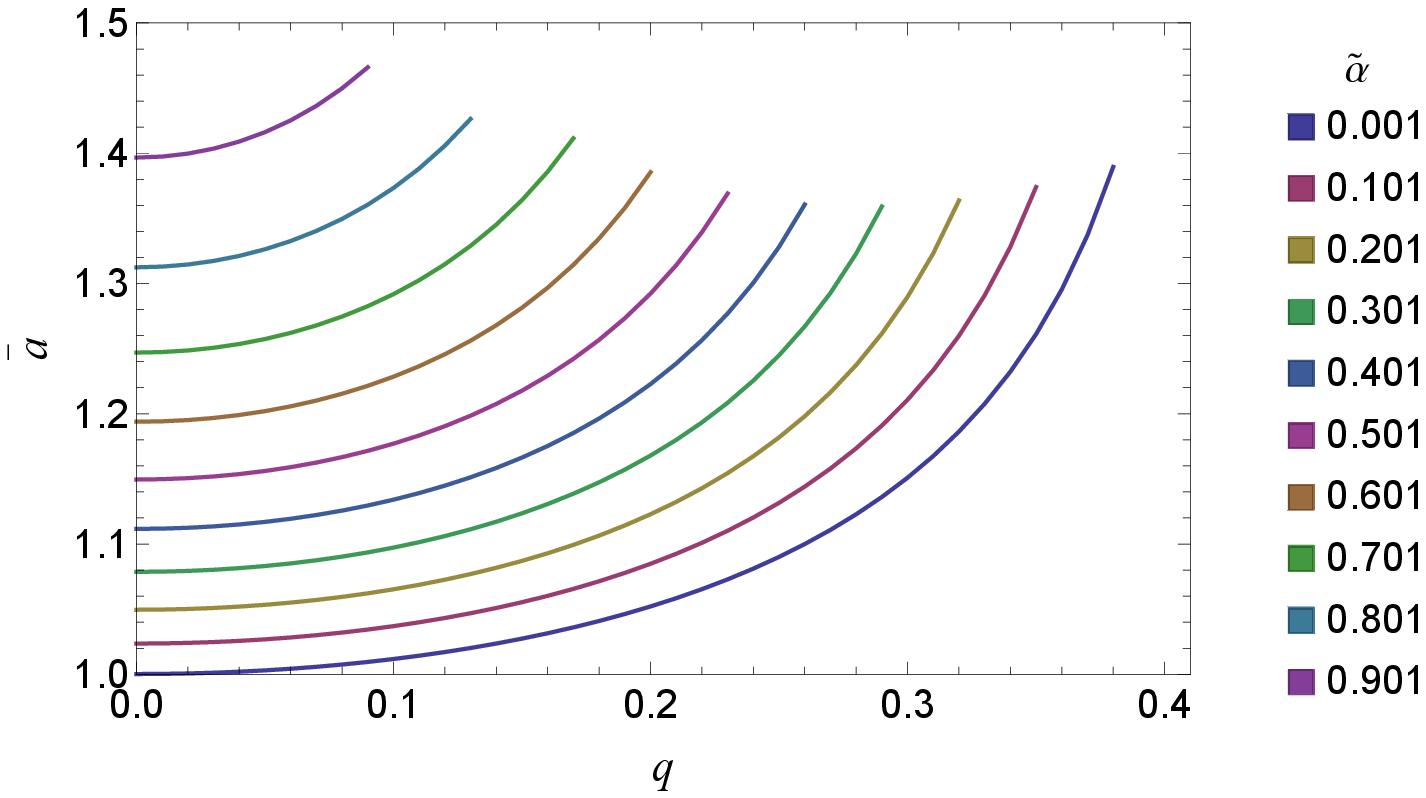}&
			\includegraphics[scale=0.62]{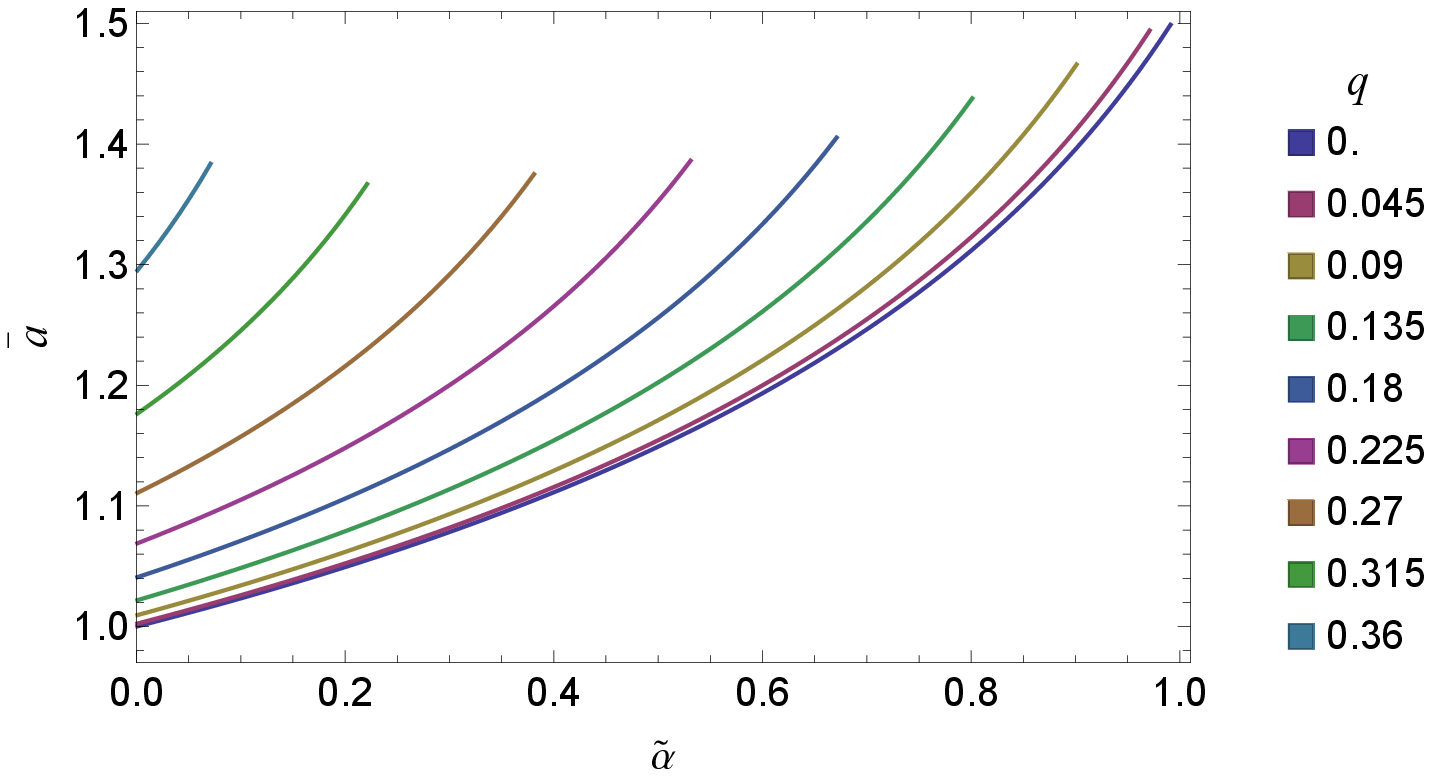}\\
			\includegraphics[scale=0.62]{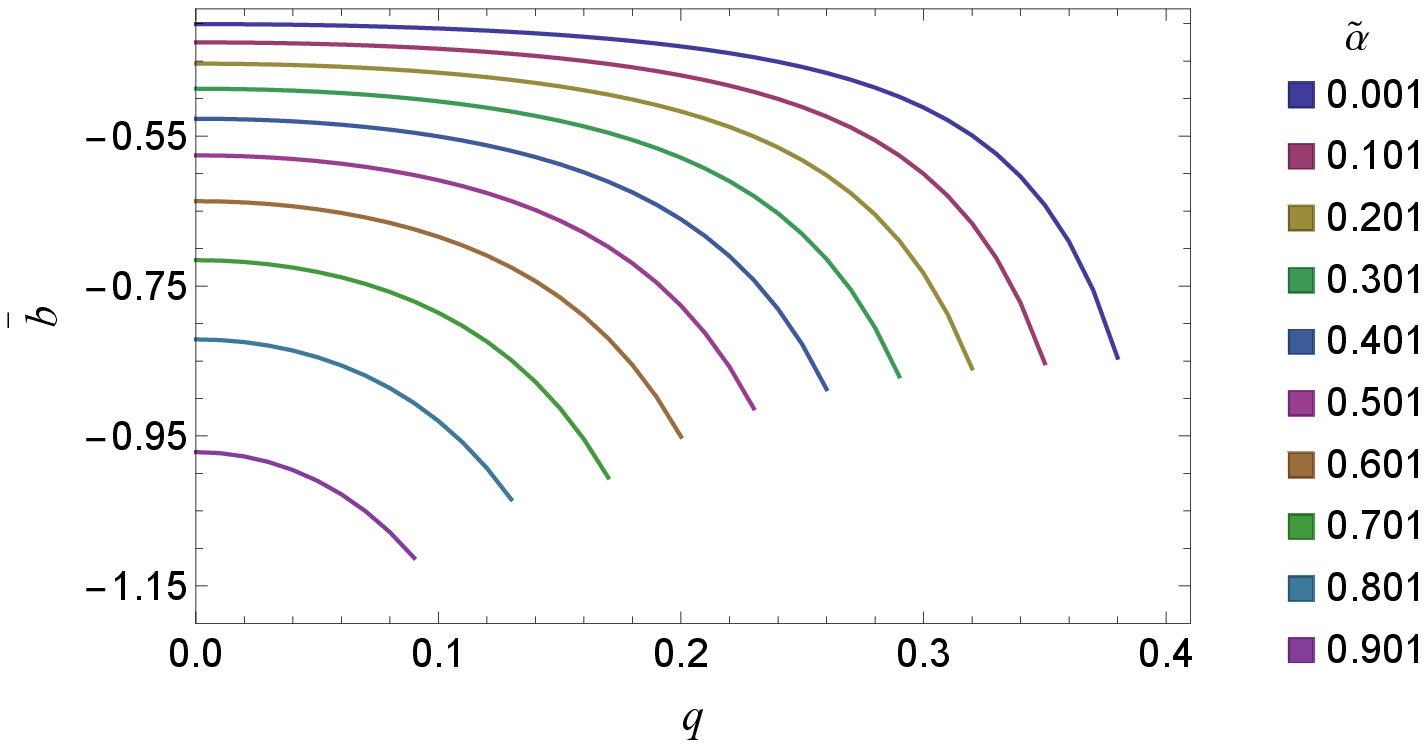}&
			\includegraphics[scale=0.62]{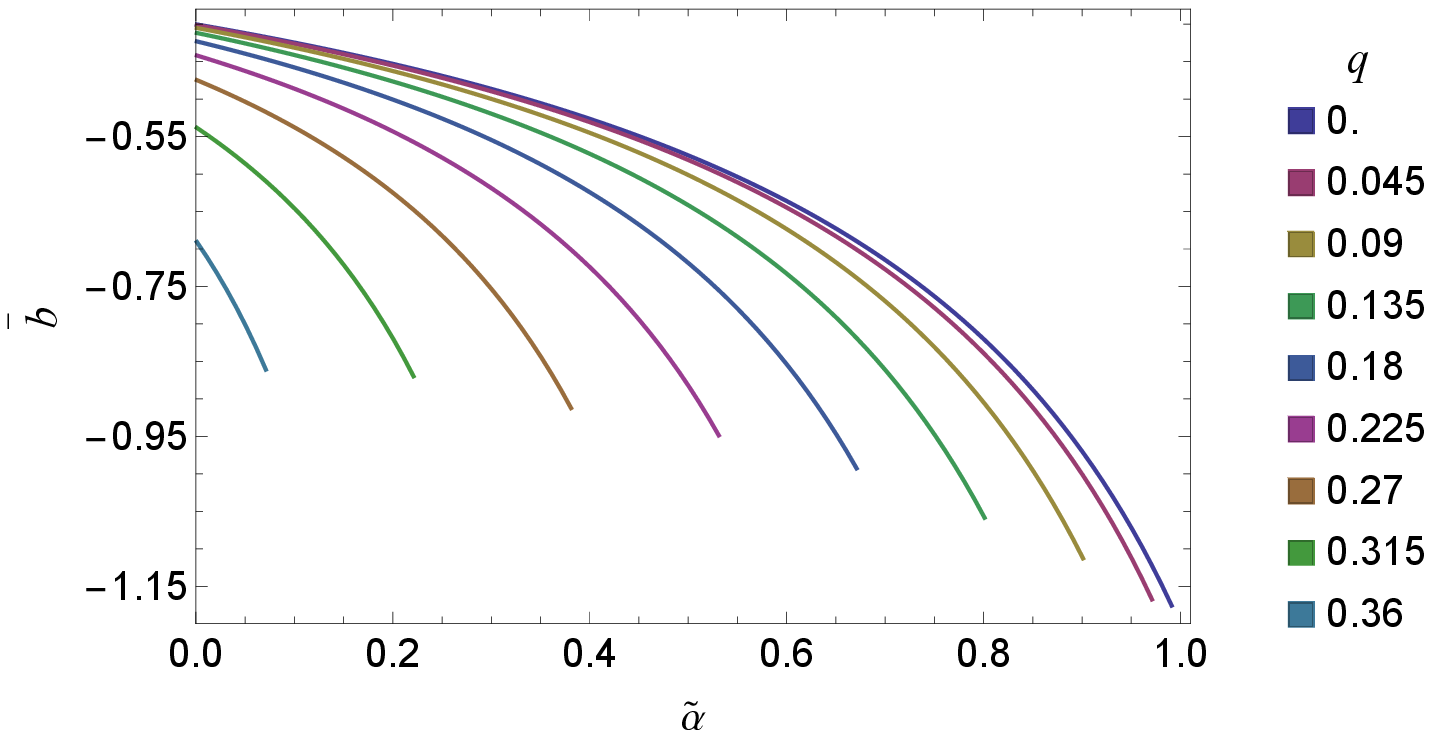}\\
		\end{tabular}
	\end{centering}
	\caption{The behavior of critical impact parameter $u_m$ (\textit{upper panel}), lensing coefficients $\bar{a}$ (\textit{middle panel}) and $\bar{b}$ (\textit{lower panel}) as a function of   $q$ for different values of  $\tilde{\alpha}$ (\textit{left}) and as a function of  $\tilde{\alpha}$ for different values of $q$ (\textit{right}). Our results in the limits $\tilde{\alpha}\to 0$ encompass those of Bardeen black holes, those of Schwarzschild black hole when $\tilde{\alpha}\to 0,~q=0$ and if $q=0$, $\tilde{\alpha}\ne 0$, we obtain  results of the 4D EGB black holes.}\label{plot6}		
\end{figure*} 

Next, $u_m=u(x_m)$ is the critical impact parameter whose  behaviour  is depicted in Fig.~\ref{plot6}.  The dependence of $u_m$ on $\tilde{\alpha}$  and  $q$ are qualitatively  the same as that of $x_m$ \citep{Eiroa:2010wm,Islam:2020xmy}. The light rays with impact parameter $u<u_m$ fall into the black hole whereas get scattered by black hole when $u>u_m$ (cf. Fig.~\ref{plot5}).  The deflection caused by the black hole increases as the impact parameter decreases (cf. Fig.~\ref{plot5}). The deflection angle for null geodesics as obtained by Virbhadra {\it et al.} \citep{Virbhadra:1998dy}
\begin{equation}\label{def}
\alpha_D(x_0) = -\pi + 2\int_{x_0}^\infty \frac{1}{\sqrt{A(x)C(x)}\sqrt{\frac{C(x)A(x_0)}{C(x_0)A(x)}-1}}dx.    
\end{equation}
\begin{figure*}
\begin{center}	
\begin{tabular}{p{9cm} p{9cm}}
\includegraphics[scale=0.62]{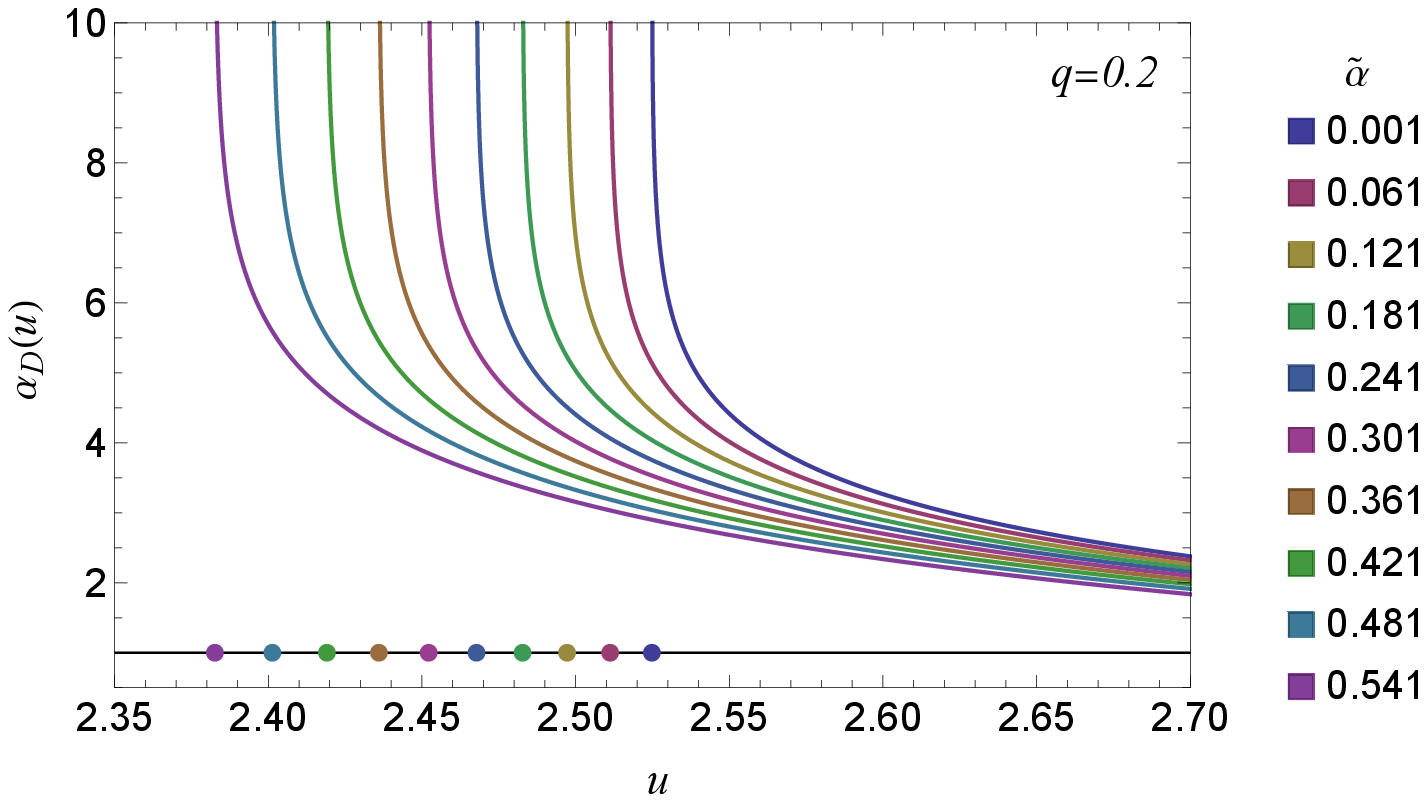}&
\includegraphics[scale=0.62]{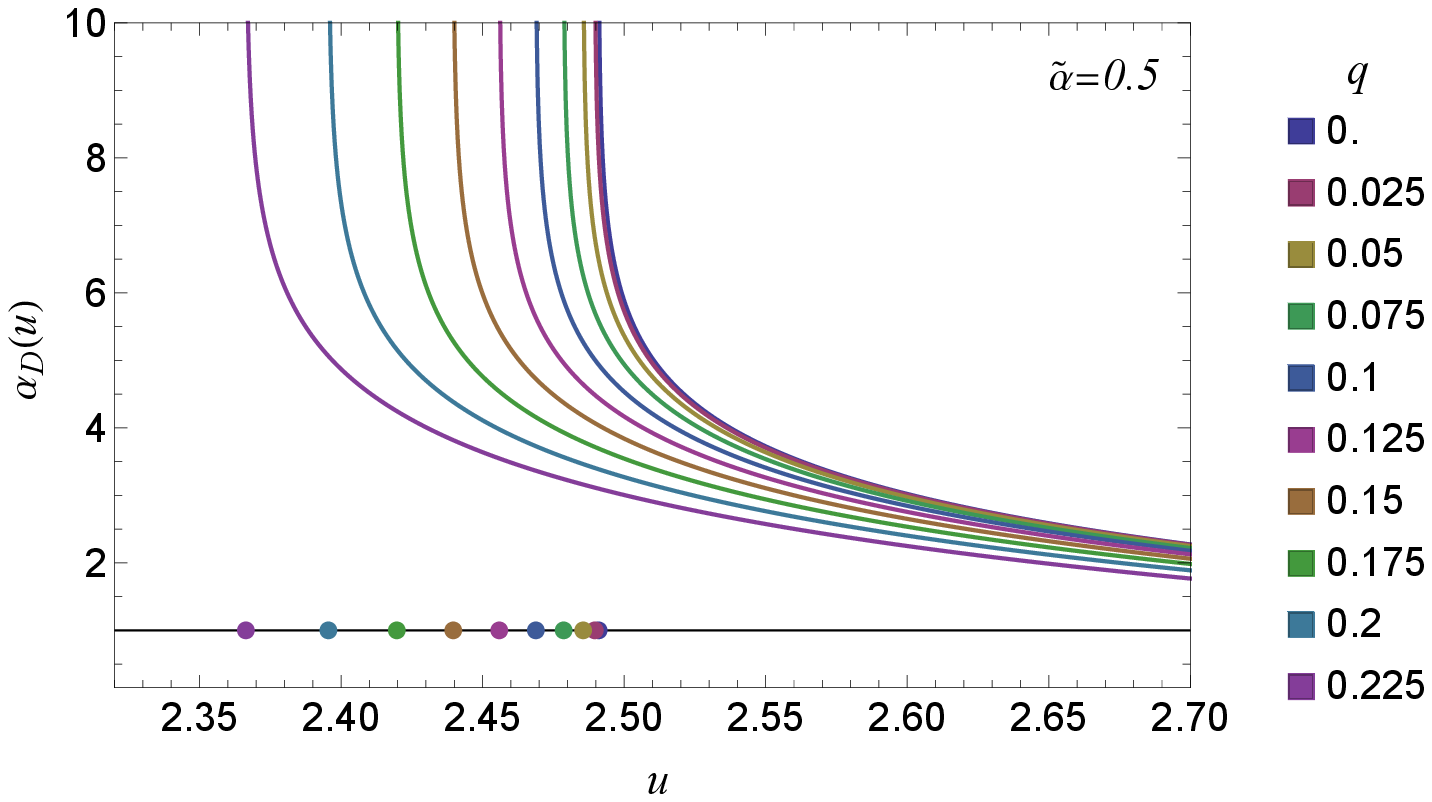}\\
\end{tabular}
\caption{The deflection angle is plotted against the impact parameter  for a given value of $q$ and different values of  $\tilde{\alpha}$ (\textit{left}) and  given value of  $\tilde{\alpha}$ for different values of $q$ (\textit{right}). The coloured bullet points on the $x-$axis correspond to the impact parameter at which deflection angle diverges. }\label{plot7}	
\end{center}   
\end{figure*}
\begin{figure*}
	\begin{centering}
		\begin{tabular}{p{9cm} p{9cm}}
		   \includegraphics[scale=0.6]{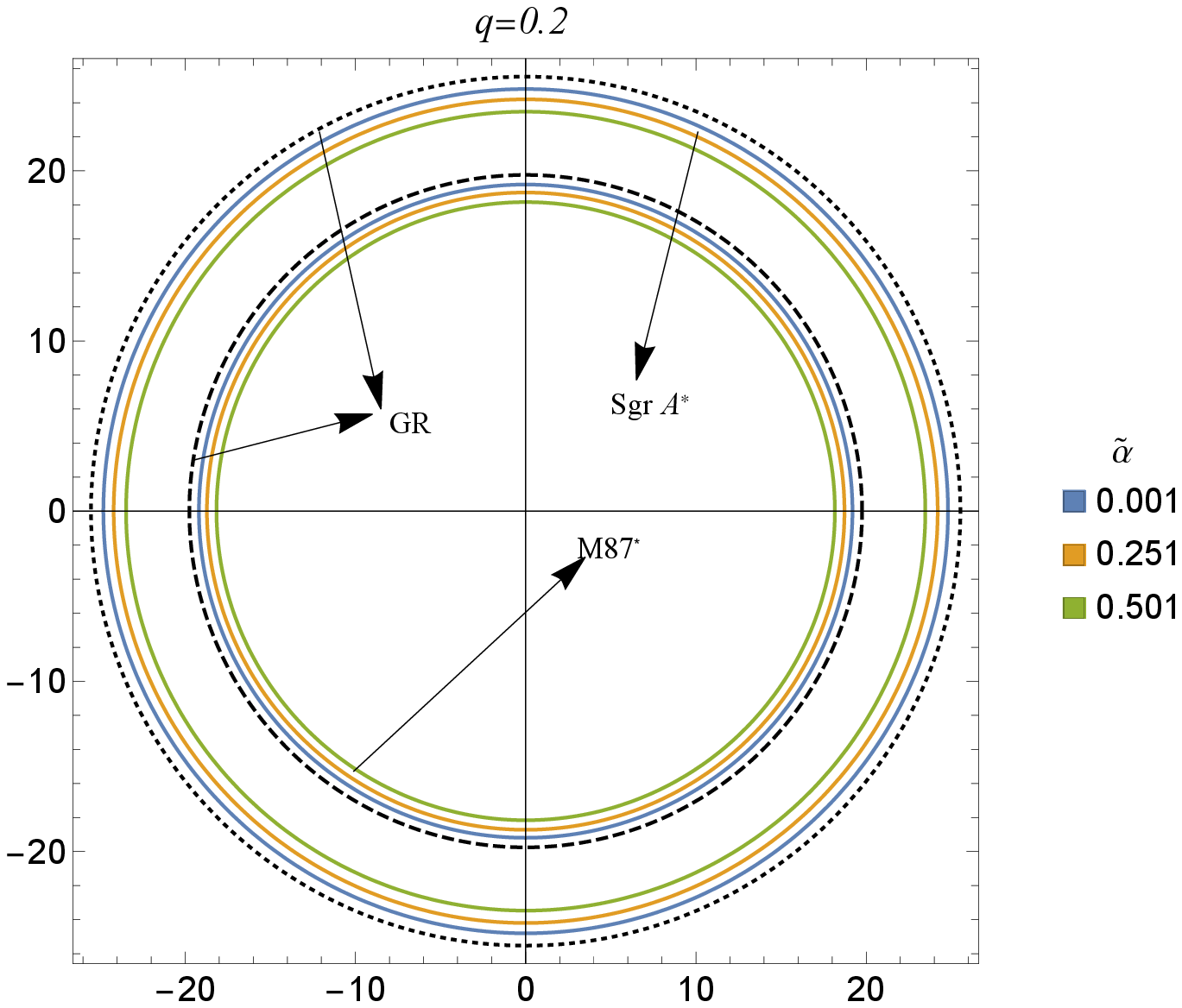}&
			\includegraphics[scale=0.6]{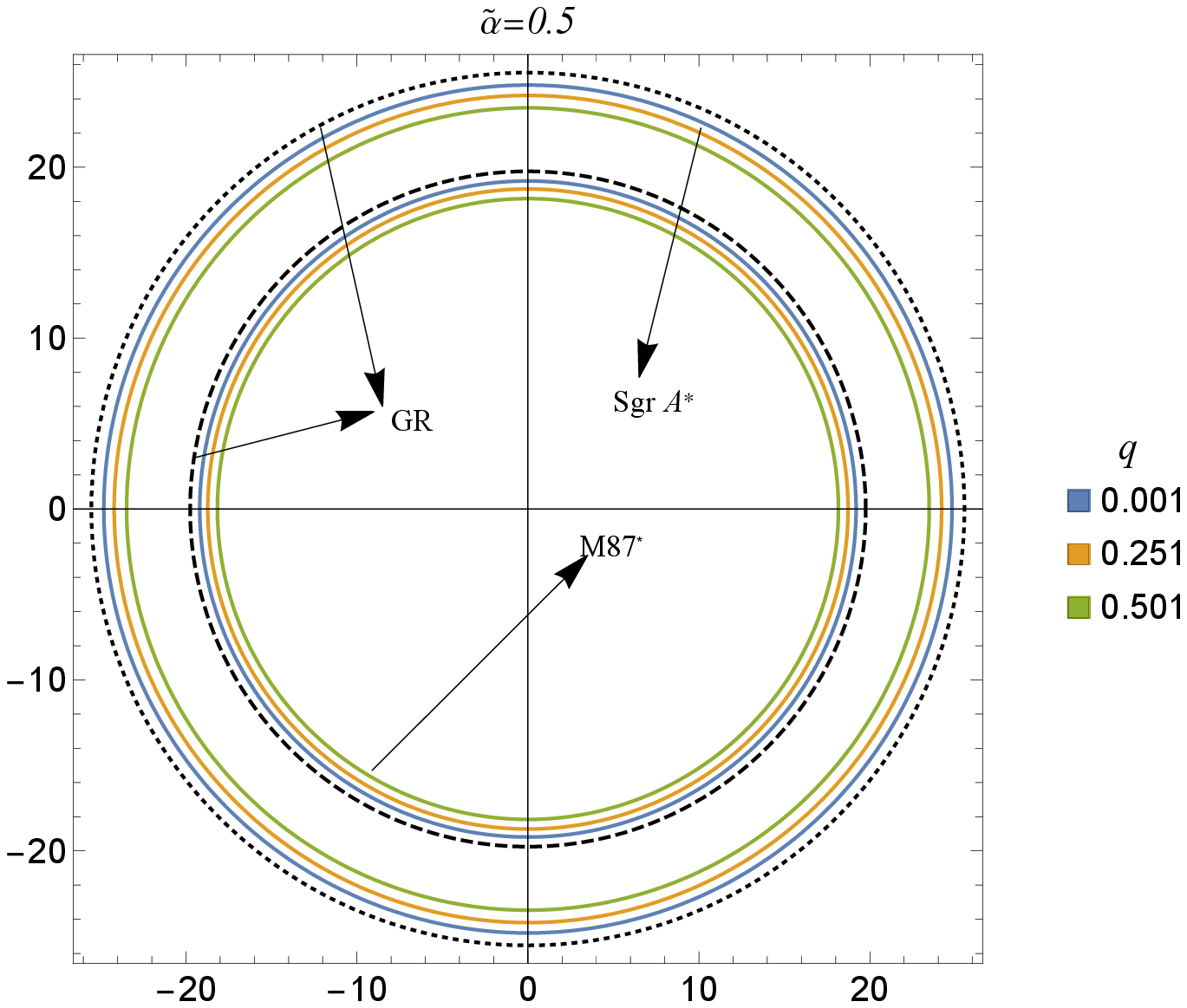}
			\end{tabular}
	\end{centering}
	\caption{Formation of outer most relativistic Einstein ring  for a given value of $q$ and different values of  $\tilde{\alpha}$ (\textit{left}) and  given value of  $\tilde{\alpha}$ for different values of $q$ (\textit{right}). The outer dotted and dashed rings, respectively, correspond  to the case when Sgr A* and M87* are considered as Schwarzschild black holes ($\tilde{\alpha}\to 0$ and  $q=0$). In the limits $\tilde{\alpha}\to 0$ and $q=0$, our results, respectively, encompass those of Bardeen black holes and 4D EGB black holes.} \label{plot8b}		
\end{figure*} 
The deflection angle increases as $x_0 \to x_m$ and diverges  at $x_0 = x_m$. We  expand the integral near the photon sphere \citep{Virbhadra:1999nm,Claudel:2000yi,Bozza:2002zj}  by  defining a new variable $z=1-x_0/x$ in SDL \citep{Tsukamoto:2016jzh,Zhang:2017vap}. This technique  not only shows the behaviour of photons near the photon sphere but also provides an analytical representation of the deflection angle, which for 4D EGB Bardeen black holes reads \citep{Bozza:2002zj,Kumar:2020sag,Islam:2020xmy}
\begin{eqnarray}\label{def4}
\alpha_{D}(u) &=& \bar{a} \log\left(\frac{u}{u_m} -1\right) + \bar{b} + \mathcal{O}(u-u_m),  
\end{eqnarray}  
where $\bar{a}$ and $\bar{b}$ are the lensing coefficients. The details of this calculation can be found in  \cite{Bozza:2002zj,Kumar:2020sag}. The parameters $\bar{a}$ and $\bar{b}$  for 4D EGB Bardeen black holes   are depicted in Fig.~\ref{plot6} and also tabulated in Table~\ref{table1}. Our results coincide with the Schwarzschild black hole of $\bar{a}=1$ and $\bar{b}=-0.4004$ at $\tilde{\alpha}\to0$ and $q=0$ \citep{Bozza:2002zj,Eiroa:2010wm,Kumar:2020sag,Islam:2020xmy}. For other values of $\tilde{\alpha}$ and $q$, $\bar{a}$ is always larger whereas $\bar{b}$ is smaller than the corresponding values of Bardeen as well as 4D EGB black holes (cf. Table~\ref{table1}).

The deflection angle is a monotonically decreasing function of  $\tilde{\alpha}$ and $q$ (cf. Fig.~\ref{plot7}) and is  sensitive to the impact parameter as $u\to u_m$ \citep{Bozza:2010xqn}, for instance, at $\tilde{\alpha}=0.5$ and $q=0.2$, $\alpha_D(\theta)= 2\pi$ (first loop) for $u=2.40566$,  which deviates from $u_m$ by $0.42\%$. Thus, depending on the impact parameter, a light ray  can make one, two  or  several  loops around the black hole   before  reaching the observer resulting  in addition to  the  primary  and  secondary  images ($|\alpha_D(\theta)|<2\pi$), two infinite sequences of relativistic images (cf.~Fig.~\ref{plot5b}), one produced by  clockwise winding of the photon and the other by 
counterclockwise winding of the photon around the black hole \citep{Virbhadra:1999nm}. These images, respectively,  are located on the same and opposite sides of the source.
\section{Image position, angular separation and magnification}\label{sect5}
We assume that the observer and  source are almost aligned along the optical axis and are placed in flat spacetime while the curvature affects the deflection angle near the lens only \citep{Bozza:2010xqn}. Further, we consider that the source is located behind the lens, and  the lens equation reads \citep{Bozza:2008ev,Islam:2020xmy}
\begin{eqnarray}\label{lenseq}
\beta &=& \theta -\frac{D_{LS}}{D_{OS}}\Delta \alpha_n,
\end{eqnarray}
where  $\Delta \alpha_n = \alpha_{D} - 2n\pi$ is the extra deflection angle with $0<\Delta \alpha_n<1$   and  $n \in N$. Also, $\theta$ and $\beta$ are the angular separation of image and source from the optic axis whereas $D_{LS}$ and $D_{OS}$, respectively, are the  distances of lens and observer from the source (cf.~Fig.~\ref{plot5b}). Using the  relation  $u=D_{OL}\tan(\theta)\approx \theta D_{OL}$,  Eq.~(\ref{def4}) together with Eq.~(\ref{lenseq}) yield  the angular position of the $n$th relativistic image $\theta_n$ in terms of lensing coefficients as \citep{Bozza:2002zj,Eiroa:2010wm,Kumar:2020sag,Islam:2020xmy}
\begin{equation}\label{theta}
\theta_n = \theta_n^0 + \frac{(D_{OL}+D_{LS})}{D_{LS}}\frac{u_m e_n}{D_{OL}\bar{a}}(\beta-\theta_n ^0),
\end{equation}     
where 
\begin{equation}
\theta_n ^0 = \frac{u_m}{D_{OL}}(1+e_n), ~~~~~ e_n = \text{exp}\left({\frac{\bar{b}}{\bar{a}}-\frac{2n\pi}{\bar{a}}}\right).  
\end{equation}

\begin{table*}
\caption{Estimates for lensing observables and  lensing coefficients for the black hole Sgr A* and M87*  for different values of $\tilde{\alpha}$ and $q$. $R_s = 2GM/c^2$ is the Schwarzschild radius. Our results in limits $\tilde{\alpha}\to 0$ encompass those of Bardeen black holes, those of Schwarzschild black hole when $\tilde{\alpha}\to 0,~q=0$ and if $q=0$, $\tilde{\alpha}\ne 0$, we obtain  results of the 4D EGB black holes.} \label{table1}  
\resizebox{1\textwidth}{!}{
 
	\begin{tabular}{p{1cm} p{1.2cm} p{1.5cm} p{2cm} p{1.5cm} p{2cm} p{1.5cm} p{1.5cm} p{1.5cm} p{1.5cm}}
		
\hline\hline
\multicolumn{2}{c}{}&
 \multicolumn{2}{c}{Sgr A*} & 
 \multicolumn{2}{c}{M87*}  & 
 \multicolumn{1}{c}{}&
 \multicolumn{3}{c}{Lensing Coefficients}\\
{$\tilde{\alpha}$ } & {$q$} & {$\theta_\infty $ ($\mu$as)} & {$s$ ($\mu$as) }  & {$\theta_\infty $ ($\mu$as)} & {$s$ ($\mu$as) } &  {$r_{\text{mag}}$} & {$\bar{a}$}&{$\bar{b}$} & {$u_m/R_s$}\\ 
\hline
\multirow{4}{*}{0.0001}   & 0.00  & 25.56427 & 0.0319976 & 19.7819 & 0.0247601 & 6.82173 & 1.00002 & -0.400252 & 2.59806 \\
            & 0.10  & 25.3917 & 0.0340962 & 19.6483 & 0.026384 & 6.74409 & 1.01153 & -0.40608 & 2.58052 \\
            & 0.20  & 24.84556 & 0.0420186 & 19.2257 & 0.0325144 & 6.48576 & 1.05182 & -0.429893 & 2.52502 \\
            & 0.36  & 22.81603  & 0.1044801 & 17.6553 & 0.0808477 & 5.27011 & 1.29445 & -0.688999 & 2.31876 \\
\hline 
\multirow{4}{*}{0.1}   & 0.00 & 25.37146 & 0.0361142 & 19.6327 & 0.0279455 & 6.66617 & 1.02336 & -0.424612 & 2.57846 \\                                  
& 0.15 & 24.96031 & 0.04240501 & 19.3145 & 0.0328134 & 6.46588 & 1.05506 & -0.445738 & 2.53668 \\
& 0.25  & 24.15324 & 0.0595154 & 18.69 & 0.0460536 & 6.0308 & 1.13117 & -0.510569 & 2.45466 \\      
& 0.35 & 22.64073 & 0.1253277 & 17.5196 & 0.0969797 & 4.96981 & 1.37266 & -0.84997 & 2.30094 \\
\hline
\multirow{4}{*}{0.5}   & 0.00 & 24.51075 & 0.0627214 & 18.9667 & 0.0485345 & 5.9365 & 1.14914 & -0.575084 & 2.49099 \\
& 0.10  & 24.29243 & 0.0798644 & 18.5756 & 0.0617999 & 5.60461 & 1.21719 & -0.661753 & 2.43964 \\
& 0.20  & 23.57138 & 0.0997613 & 18.2398 & 0.0771963 & 5.28221 & 1.29148 & -0.774778 & 2.39552 \\
& 0.23 & 23.22022 & 0.1207592 & 17.968 & 0.0934446 & 4.98679 & 1.36799 & -0.911054 & 2.35983 \\
\hline
\multirow{4}{*}{0.9}   & 0.0 & 23.4261 & 0.1296815 & 18.1273 & 0.100349 & 4.88771 & 1.39572 & -0.969719 & 2.38076 \\
& 0.04  & 23.37983 & 0.1331366 & 18.0915 & 0.103022 & 4.84519 & 1.40797 & -0.993561 & 2.37606 \\

& 0.06 & 23.32114 & 0.1376848 & 18.0461 & 0.106542 & 4.79013 & 1.42415 & -1.02578 & 2.37009 \\

& 0.09 & 23.1853 & 0.1489592 & 17.941 & 0.115266 & 4.65751 & 1.46471 & -1.11007 & 2.35629 \\     
\hline\hline
\end{tabular}
}	
\end{table*}

\begin{figure*}
	\begin{centering}
		\begin{tabular}{p{9cm} p{9cm}}
		    \includegraphics[scale=0.62]{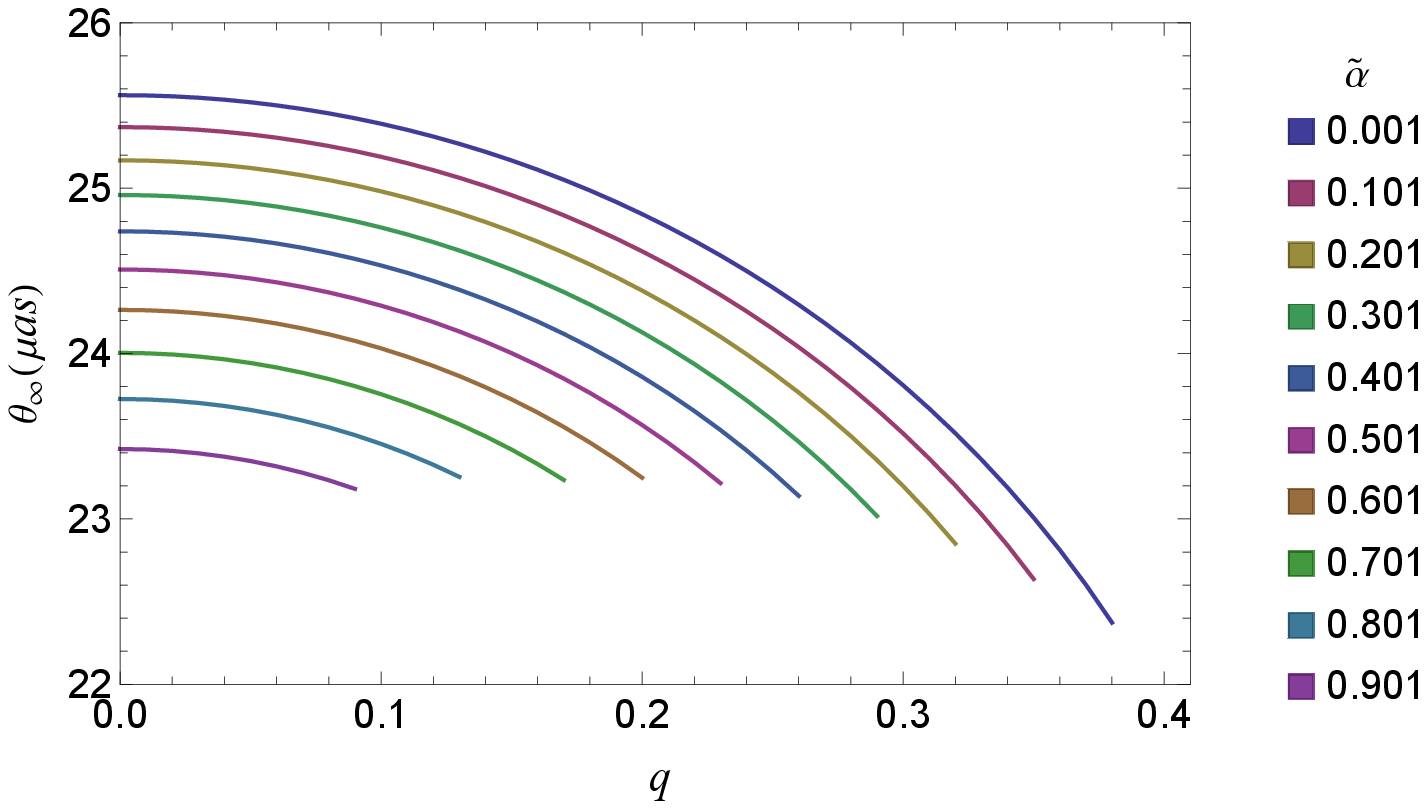}&
			\includegraphics[scale=0.62]{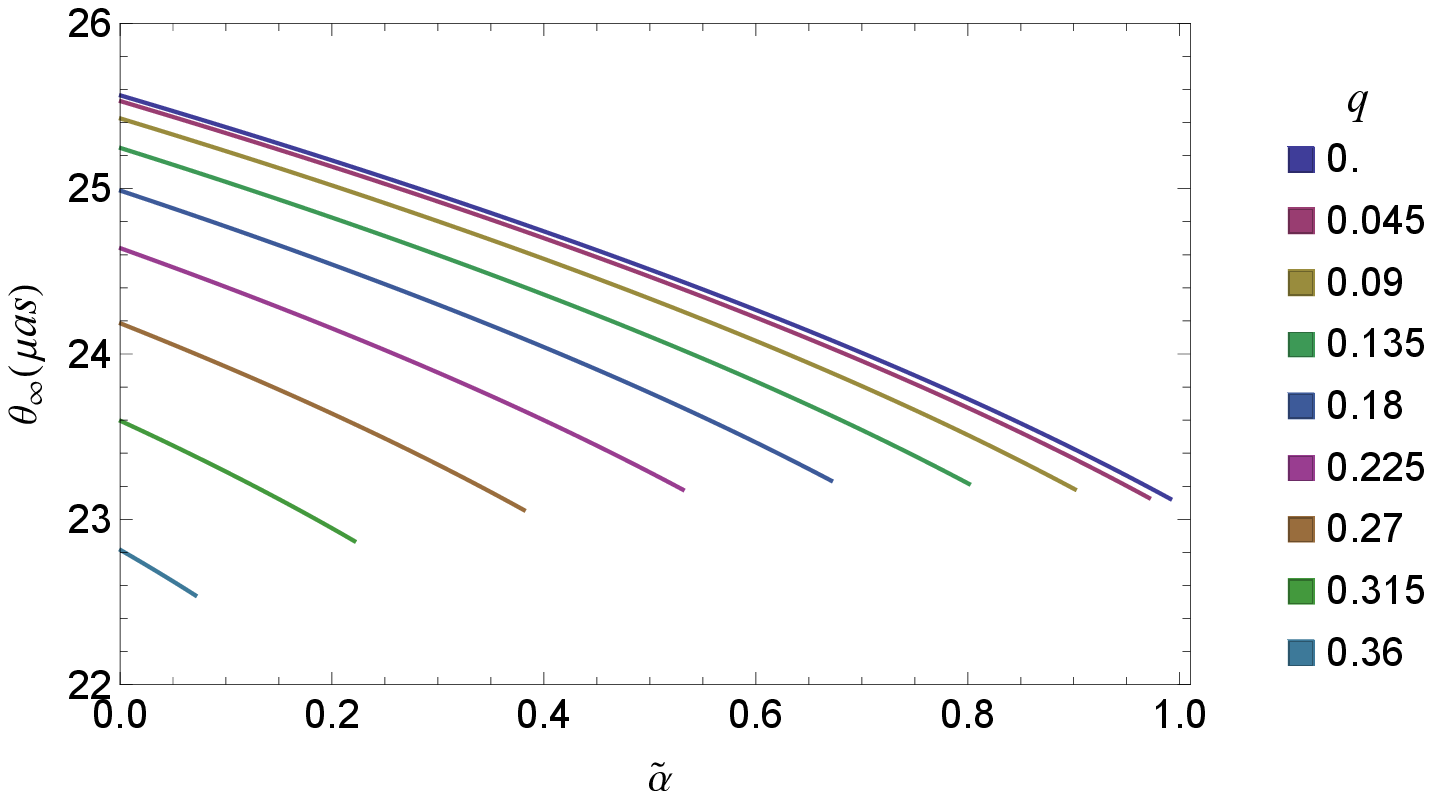}\\
			\includegraphics[scale=0.62]{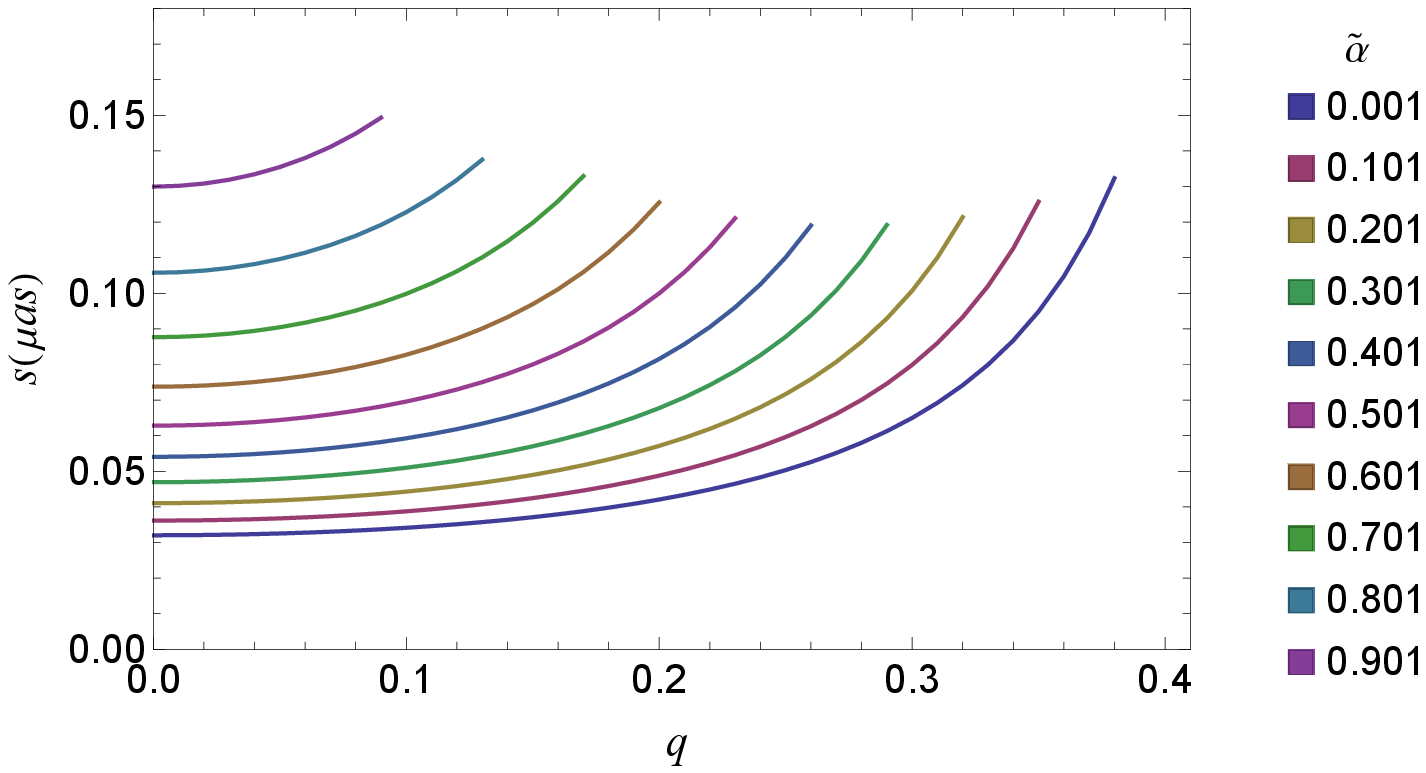}&
			\includegraphics[scale=0.62]{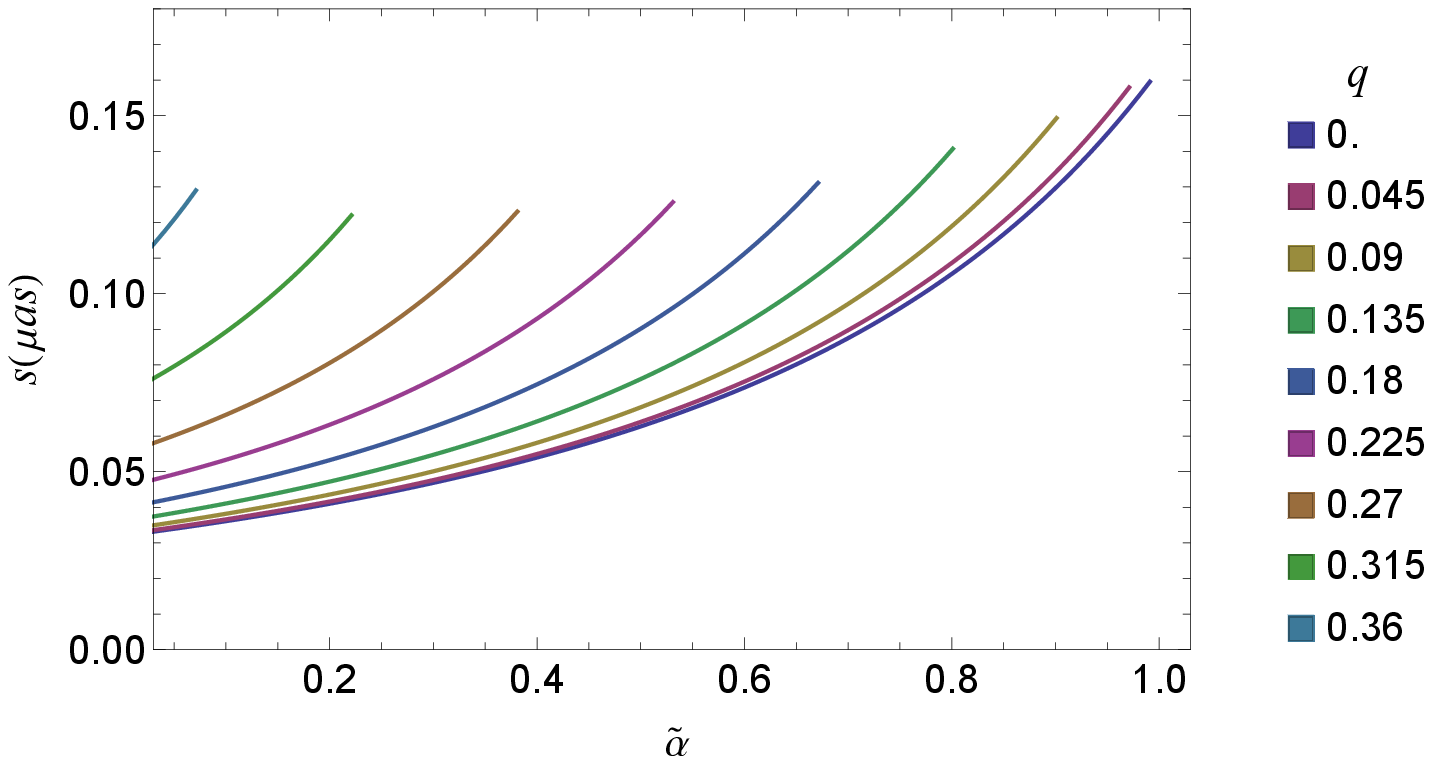}\\
		\end{tabular}
	\end{centering}
	\caption{The behavior of  lensing observables $\theta_{\infty}$ (\textit{upper panel}) and $s$ (\textit{lower panel}) as a function of   $q$ for different values of  $\tilde{\alpha}$ (\textit{left}) and as a function of  $\tilde{\alpha}$ for different values of $q$ (\textit{right}) for SgrA*. Our results in limits $\tilde{\alpha}\to 0$ encompass those of Bardeen black holes, those of Schwarzschild black hole when $\tilde{\alpha}\to 0,~q=0$ and if $q=0$, $\tilde{\alpha}\ne 0$, we obtain  results of the 4D EGB black holes.}\label{plot8}		
\end{figure*}  
The quantity $\theta_n^0$ corresponds to the value of $\theta$ when photon travels $2n\pi$ around the black hole and the second term are the corrections to $\theta_n^0$ \citep{Chen:2009eu}. In the limit $n \to \infty$, we find that $e_n\to 0$ such that $u_m=\theta_{\infty} D_{OL}$. Moreover, the case $\beta=0$ corresponds to  perfect alignment \citep{Bisnovatyi-Kogan:2017kii} and taking  $D_{OS}=2D_{OL}$ with $D_{OL}\gg u_m$, Eq.~(\ref{theta}) reduces to \citep{Eiroa:2005ag}
\begin{equation}\label{EinsteinR}
\theta_n^{E} = \frac{u_m}{D_{OL}}(1+e_n),
\end{equation}
which is the angular radius of the $n$-th relativistic Einstein ring. Note that $n=1$ corresponds to the outermost ring. The size of the ring decreases with $\tilde{\alpha}$ and $q$ (cf. Fig.~\ref{plot8b}) and is  smaller than the Bardeen as well as 4D EGB black holes. $\theta_1^{E}$ takes the maximum value in the limit $\tilde{\alpha}\to0$ and $q=0$.

Another important observable is the  magnification of the image which is the ratio of  the solid angle onto the observer subtended by the image to solid angle by the source i.e., $\mu=\text{sin}\theta d\theta/ \text{sin}\beta d\beta$. Using Eq.~(\ref{theta}), we deduce the  magnification of $n$-loop images as \citep{Bozza:2002zj,Chakraborty:2016lxo} 
\begin{eqnarray}\label{mag3}
\mu_n &=& \frac{1}{\beta} \Bigg[\frac{u_m}{D_{OL}}(1+e_n) \Bigg(\frac{D_{OS}}{D_{LS}}\frac{u_me_n}{D_{OL}\bar{a}}  \Bigg)\Bigg].
\end{eqnarray}
\begin{figure*}
	\begin{centering}
		\begin{tabular}{p{9cm} p{9cm}}
		    \includegraphics[scale=0.62]{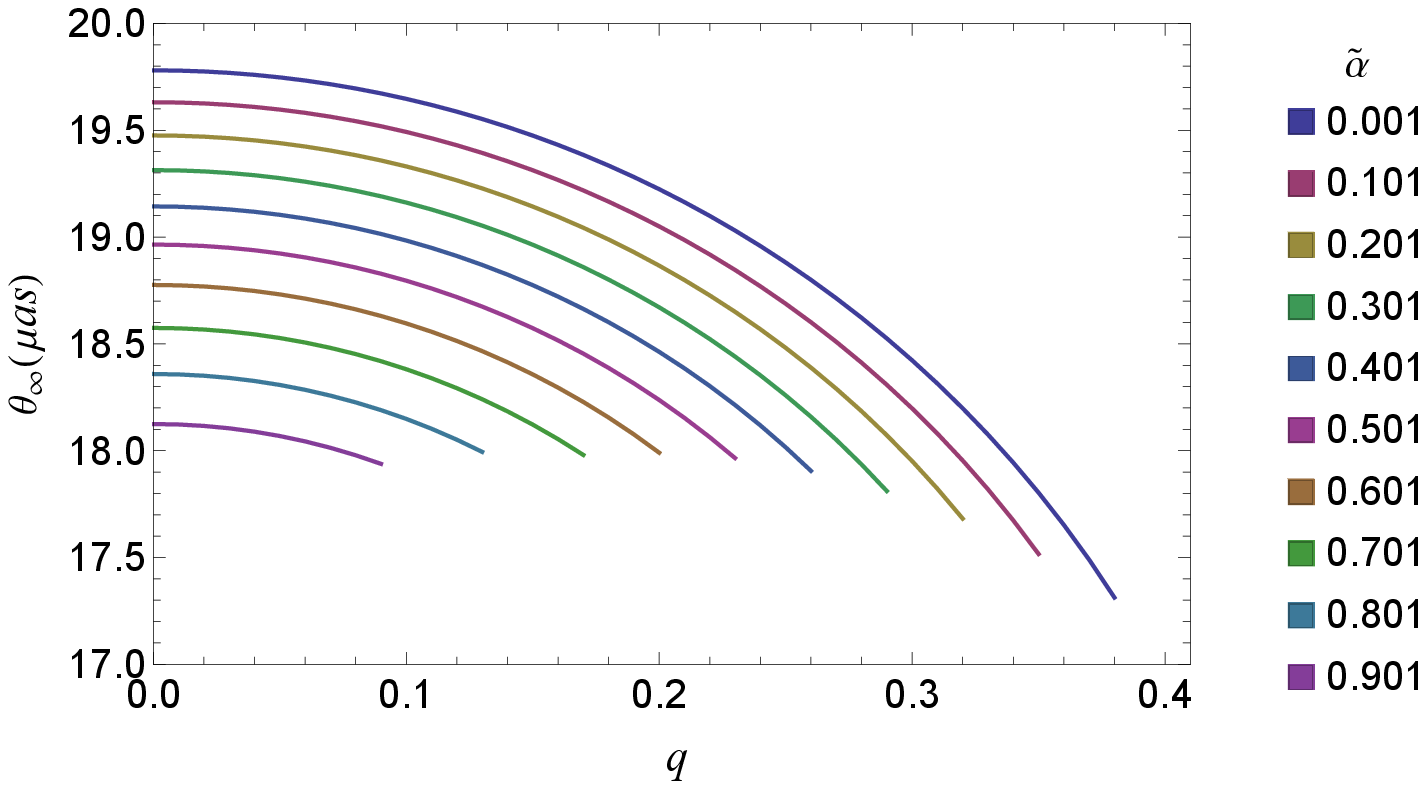}&
			\includegraphics[scale=0.62]{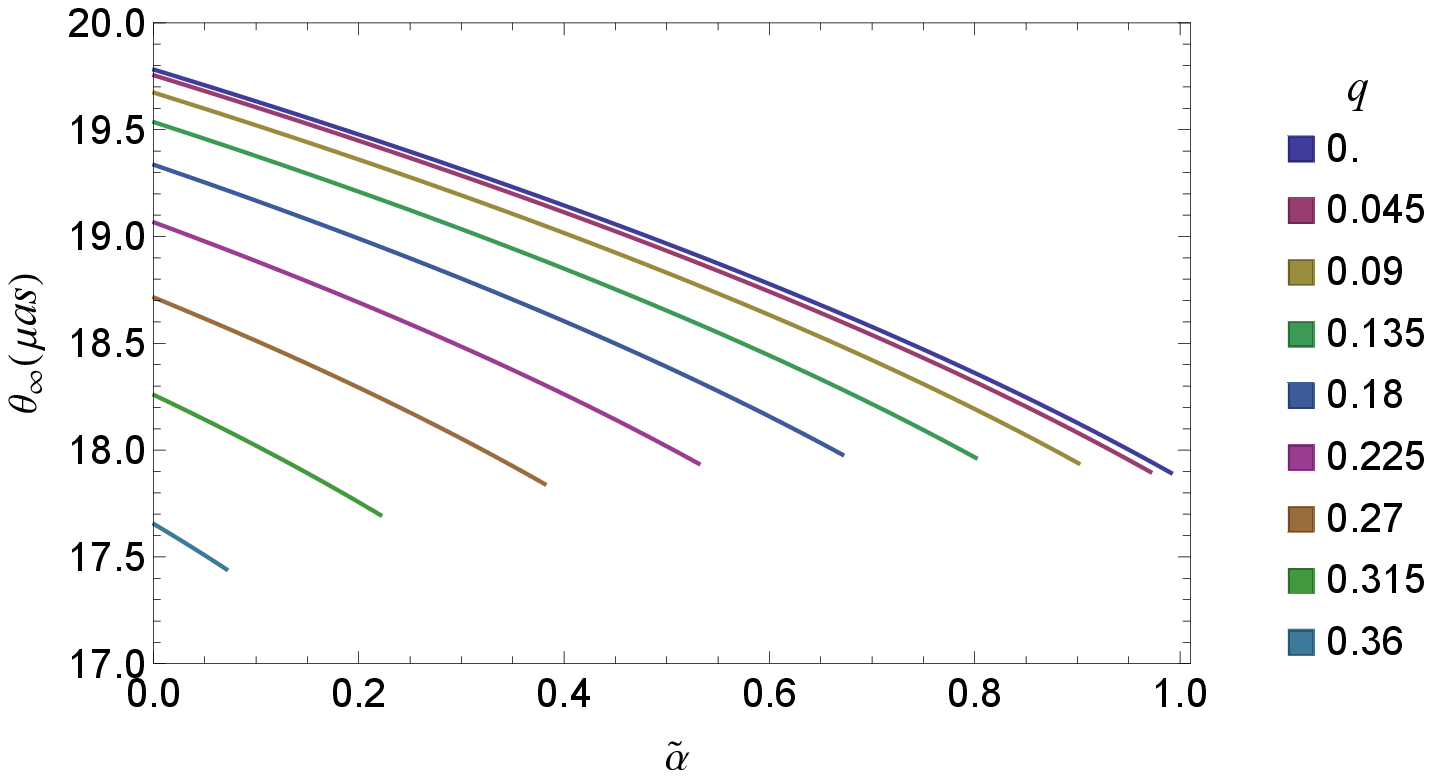}\\
			\includegraphics[scale=0.62]{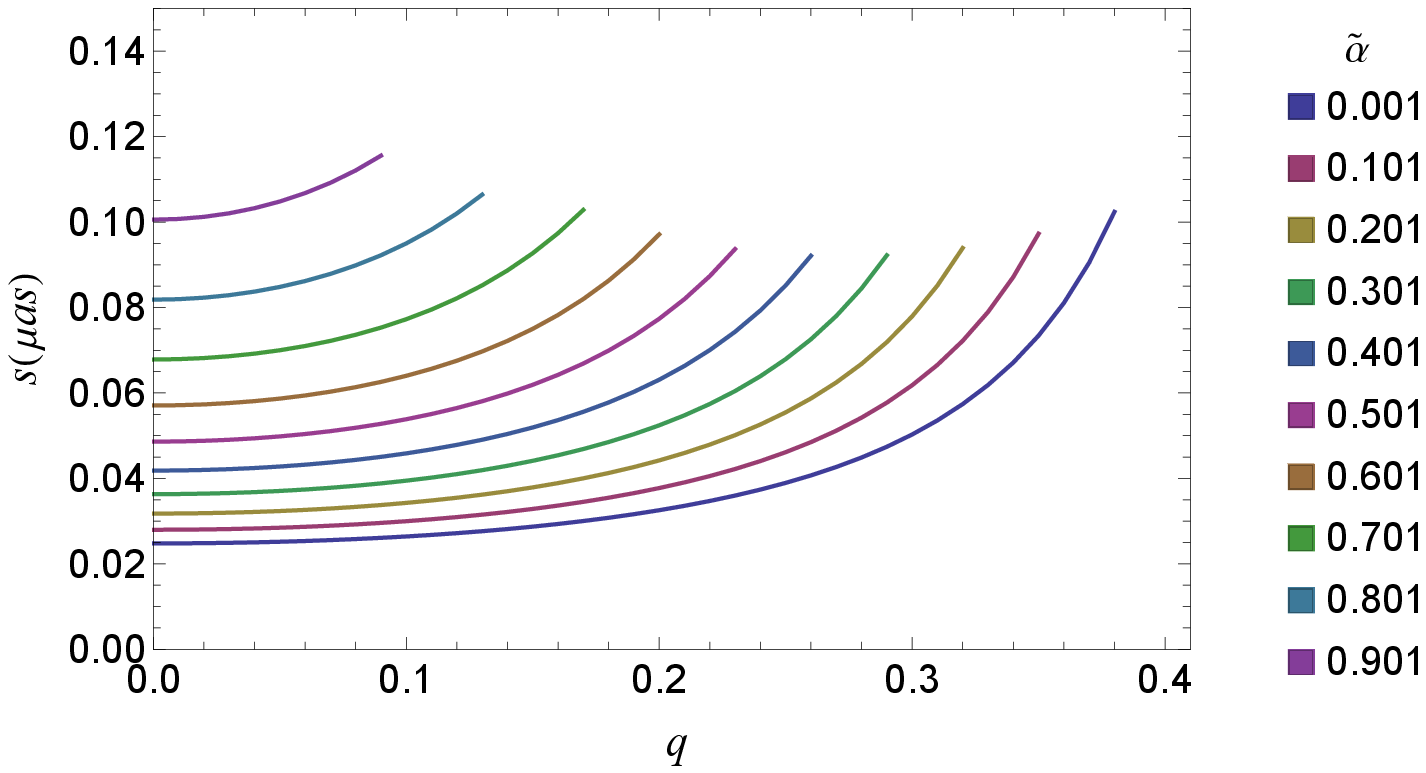}&
			\includegraphics[scale=0.62]{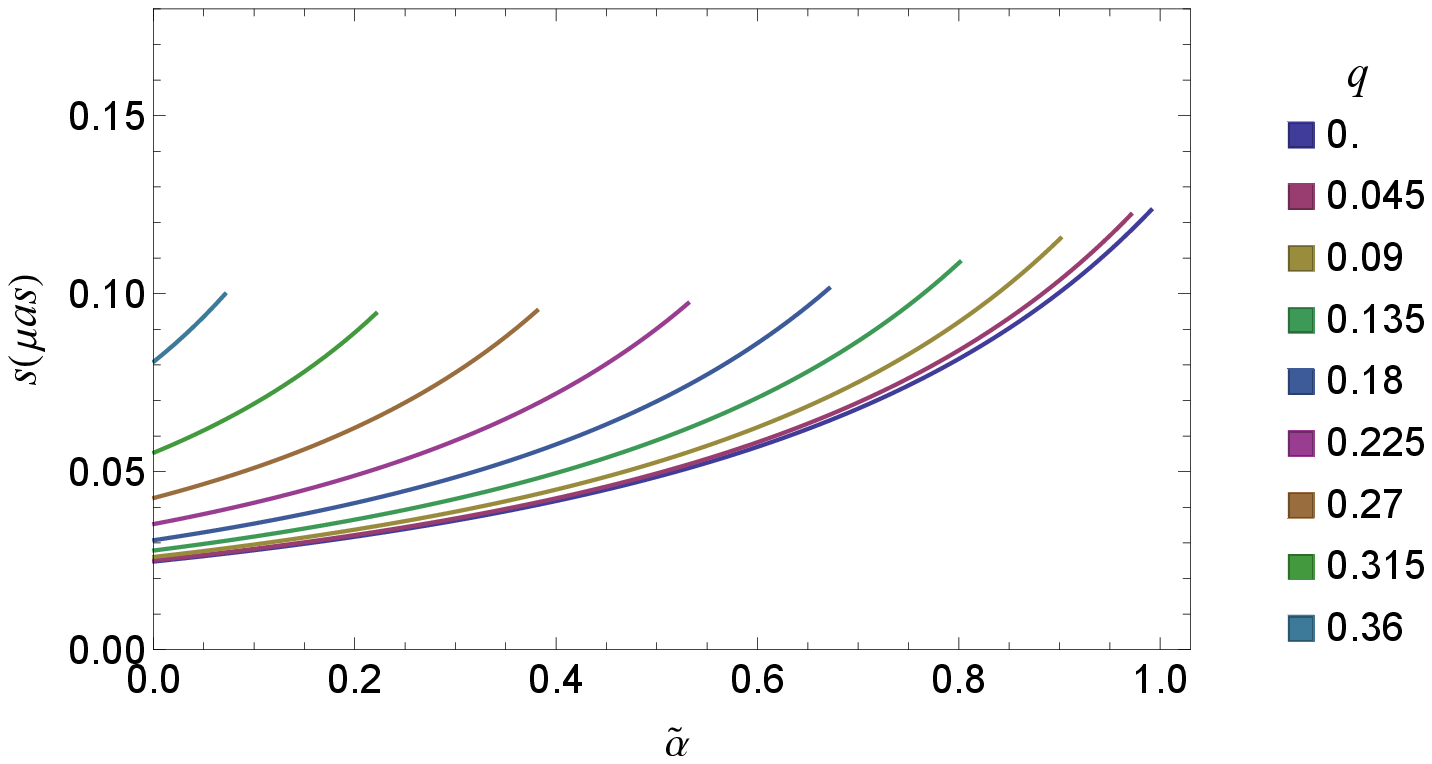}\\
		\end{tabular}
	\end{centering}
	\caption{The behavior of  lensing observables $\theta_{\infty}$ (\textit{upper panel}) and $s$ (\textit{lower panel}) with  $q$ for different values of  $\tilde{\alpha}$ (\textit{left}) and  with $\tilde{\alpha}$ for different values of $q$ (\textit{right}) for  M87*. Our results in limits $\tilde{\alpha}\to 0$ encompass those of Bardeen black holes, those of Schwarzschild black hole when $\tilde{\alpha}\to 0,~q=0$ and if $q=0$, $\tilde{\alpha}\ne 0$, we obtain  results of the 4D EGB black holes.}\label{plot9}		
\end{figure*}  
\begin{figure*}
	\begin{centering}
		\begin{tabular}{p{9cm} p{9cm}}
		   \includegraphics[scale=0.62]{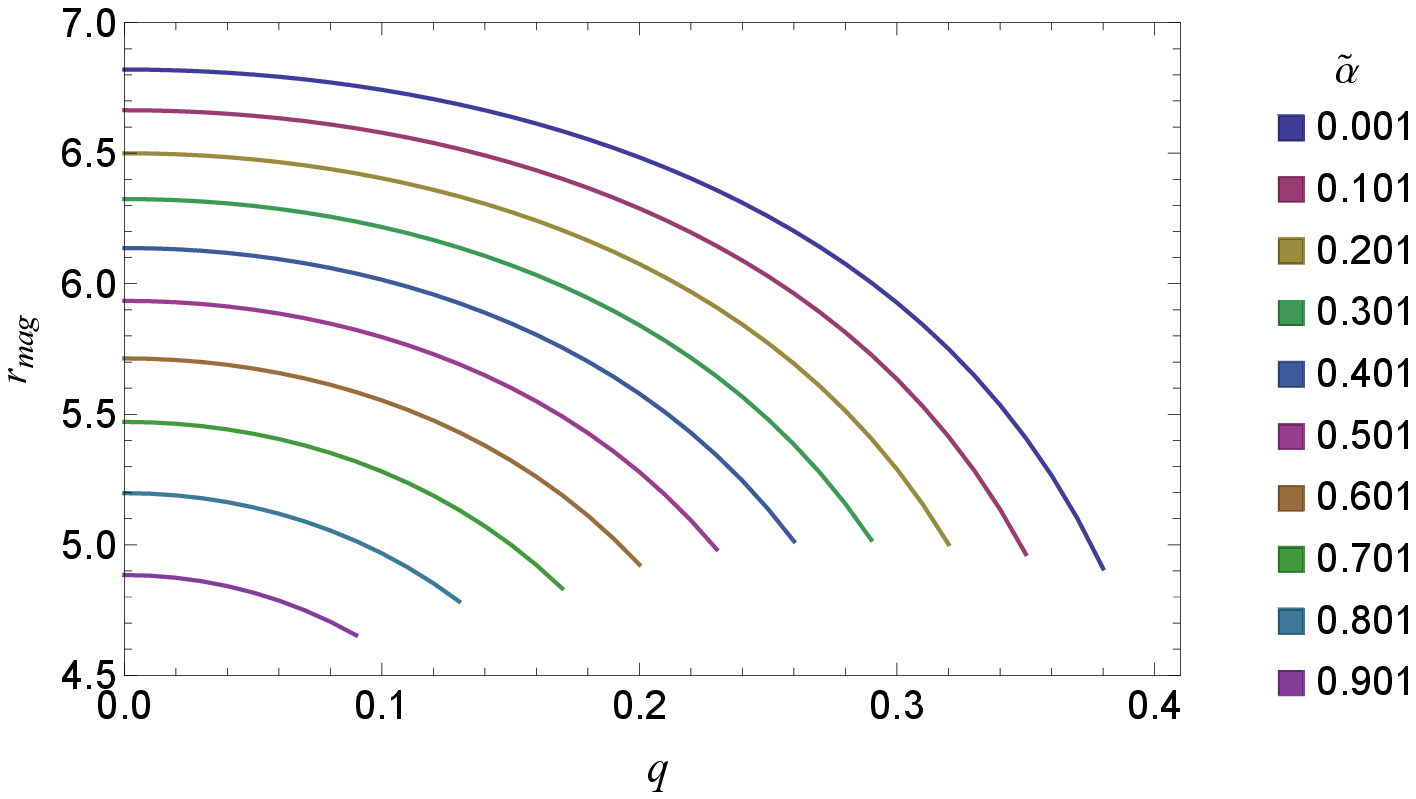}&
			\includegraphics[scale=0.62]{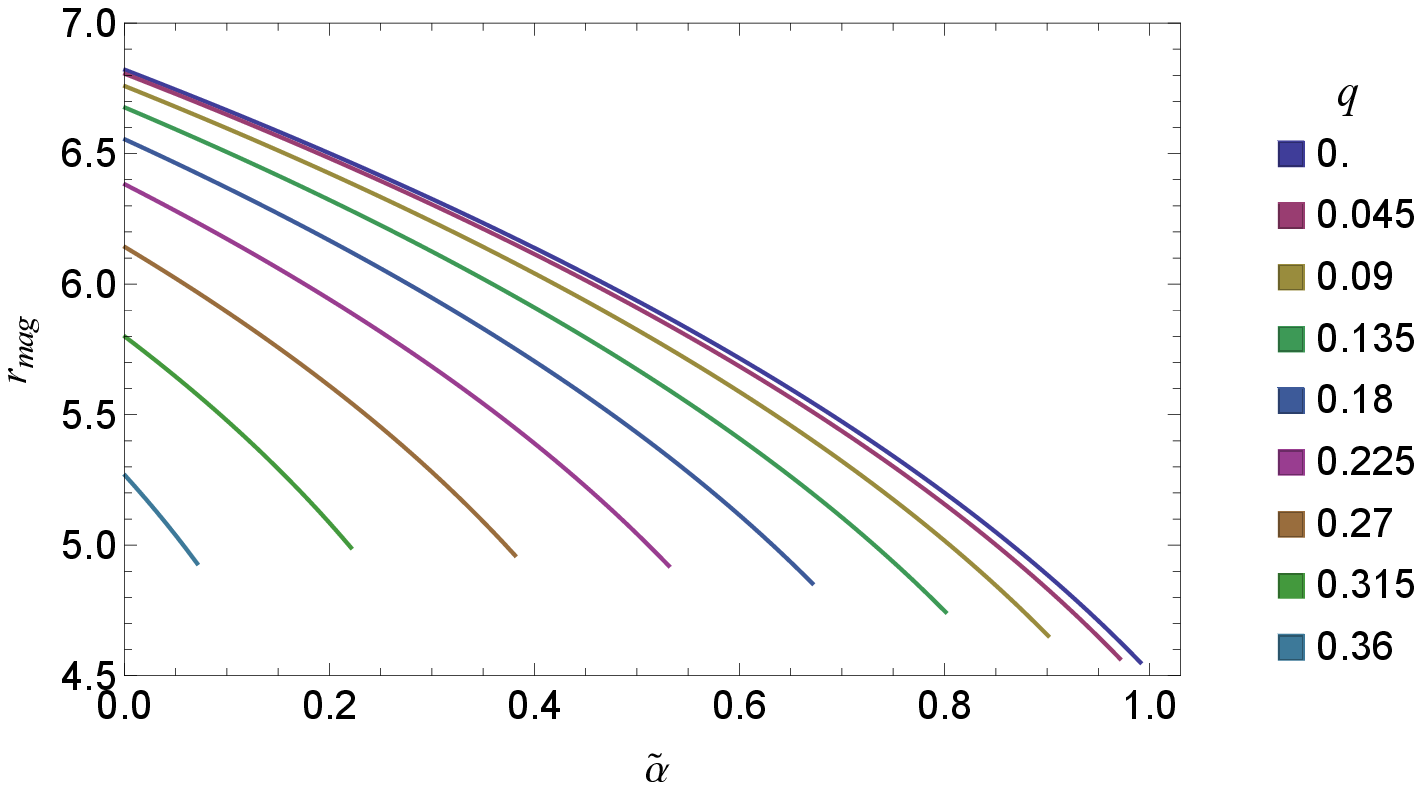}
			\end{tabular}
	\end{centering}
	\caption{The behavior of  lensing observable  $r_{mag}$ as function of   $q$ for different values of  $\tilde{\alpha}$ (\textit{left}) and  as a function of $\tilde{\alpha}$ for different values of $q$ (\textit{right}). Our results in limits $\tilde{\alpha}\to 0$ encompass those of Bardeen black holes, those of Schwarzschild black hole when $\tilde{\alpha}\to 0,~q=0$ and if $q=0$, $\tilde{\alpha}\ne 0$, we obtain  results of the 4D EGB black holes.} \label{plot10}		
\end{figure*} 
Clearly, in the limit $\beta \to 0$, $\mu_n \to \infty$ implying that the maginification of the images is maximum in the case of   perfect alignment. The Eq.~(\ref{mag3}) relates the magnification and angular position of the source to  the lensing  coefficients. As can be seen the magnification decreases with $n$ resulting in  higher order images becoming less visible.

Finally, in order to obtain the lensing coefficients, we consider the case where $\theta_1$ can be separated as a single image and remaining images are packed together at $\theta_{\infty}$. Then, we can define  three observable characteristics as \citep{Bozza:2002zj,Kumar:2020sag,Kumar:2021cyl}
\begin{eqnarray}
\theta_\infty &=& \frac{u_m}{D_{OL}},\\
s &=& \theta_1-\theta_\infty \approx  \theta_\infty~ \text{exp}\left({\frac{\bar{b}}{\bar{a}}-\frac{2\pi}{\bar{a}}}\right),\\
r_{\text{mag}} &=& \frac{\mu_1}{\sum{_{n=2}^\infty}\mu_n } = \text{exp}\left({\frac{2 \pi}{\bar{a}}}\right).
\end{eqnarray}
Here $\theta_\infty$ is the angular position acquired by the set of images in the limit $n \to \infty $ or angular radius of photon sphere, $\theta_1$ is the angular position of the outermost image, $s$ is the angular separation between the outermost image ($n=1$) and the innermost  image ($n=\infty$), and $r_{\text{mag}}$ is the magnitude of ratio  of flux between the outermost image and remaining images.
\begin{figure*}
	\begin{centering}
		\begin{tabular}{c c}
		    \includegraphics[scale=0.75]{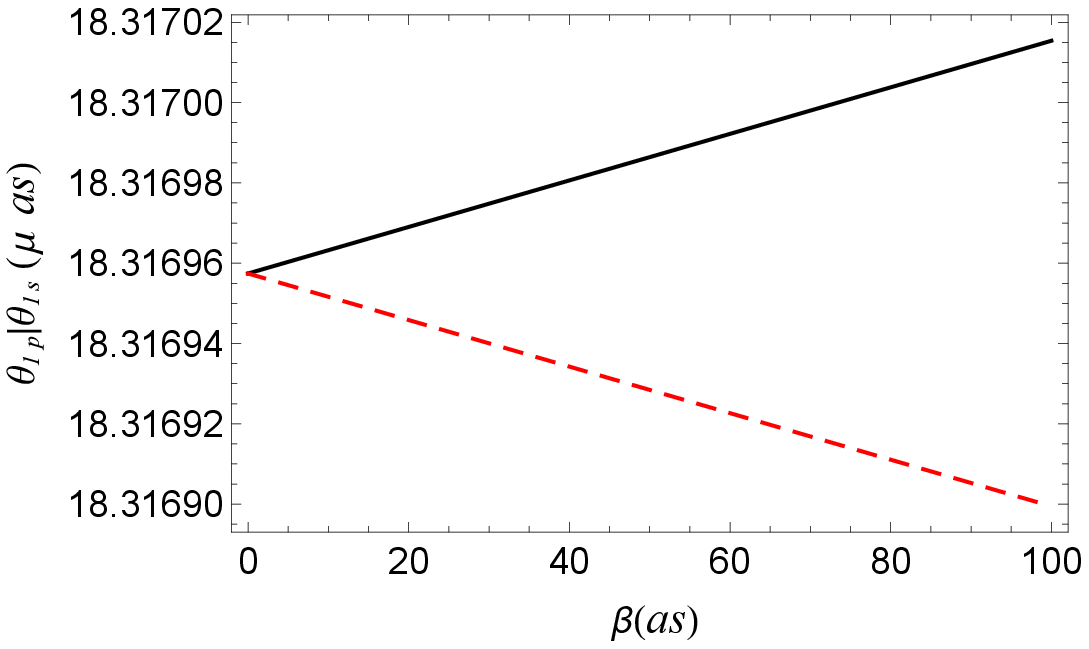}&
			\includegraphics[scale=0.75]{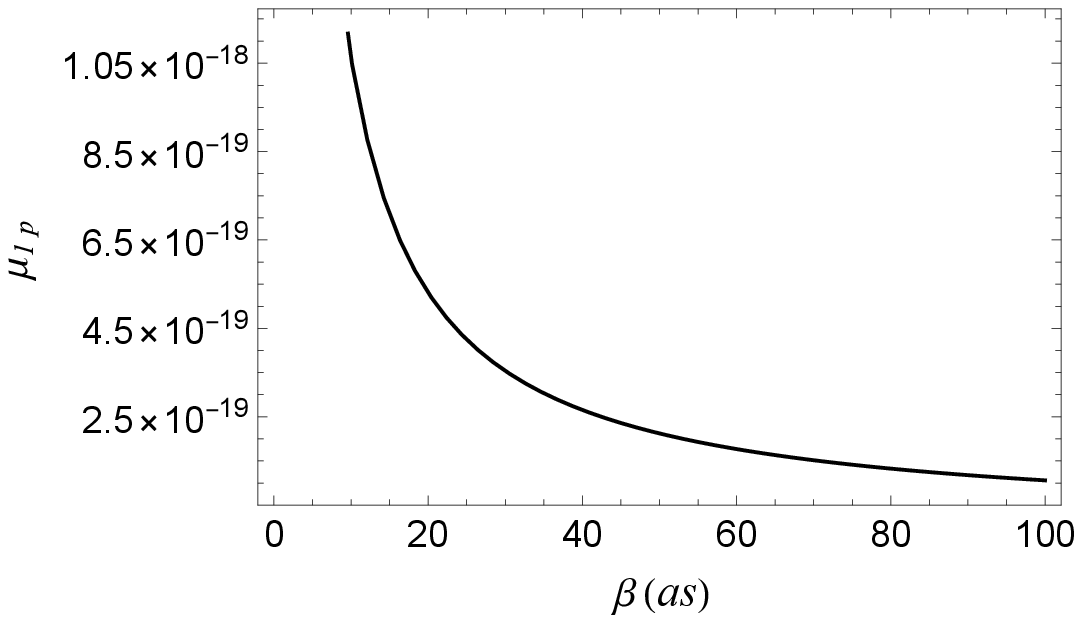}\\
			\end{tabular}
	\end{centering}
	\caption{The behavior of position of first primary $\theta_{1p}$ (solid black line) and secondary image $\theta_{1s}$ (dashed red line) with position of the source ($\beta$) for M87* (\textit{Left}). The absolute magnification of first primary image is plotted against source position for $D_{OS}=2D_{OL}$ (\textit{Right}).} \label{plot11}		
\end{figure*} 
\section{Time delay in SDL}\label{td}
Further, we derive the time delay between different relativistic images following the method developed by  \cite{Bozza:2003cp}. Time difference is caused by the fact that the photon takes different paths while winding the black hole so there is a time delay between different images which generally  depends upon which side of the lens  the images are formed.    

The images are highly demagnified and the separation between the images is of the order of $\mu$as, so we must at least distinguish the outermost relativistic image from the rest. We assume the source to be variable, which generally are abundant in all galaxies, otherwise there is no time delay to measure. Then for spherically symmetric black holes the time delay between the first and second relativistic image,  when the  two images are on the same side of the source, is given by \citep{Bozza:2003cp,Wang:2019cuf}
\begin{eqnarray}\label{deltaT}
\Delta T^s_{2,1} &=& 2\pi u_m = 2\pi D_{OL} \theta_{\infty}.
\end{eqnarray}  
We calculated the time delay for Sgr A*, M87*, and 19 other supermassive black holes \citep{Virbhadra:2007kw,Poshteh:2018wqy,Khodabakhshi:2020ddv}. Our aim is to compare the time delays between the first and second relativistic image  by 4D EGB Bardeen black holes with those of GR. Using the metric of 4D EGB and GR, we have tabulated the numerical results in  Table \ref{Table2}. For Sgr A* and M87*, the time delay can reach $\sim9.860882$~min and $\sim16023.93$~min \citep{Virbhadra:2008ws} at $\tilde{\alpha} = 0.5$ and $q=0.2$ and hence deviate from the  Bardeen black holes ($\tilde{\alpha}\to 0,~q=0.3$) by $\sim0.1001$~min and $\sim162.69$~min, respectively. If the black hole is considered a 4D EGB black hole with $\tilde{\alpha}=0.95$, the deviation from 4D EGB Bardeen black hole ($\tilde{\alpha} = 0.5,~q=0.2$)  can reach up to $\sim0.128601$~min and $\sim208.976$~min. Furthermore, the deviation from Schwarzschild black holes, respectively,  is $\sim0.8337$~min and $\sim1354.775$~min. These deviations are insignificant for Sgr A*  but for M87* and some other black holes these  are sufficiently large values  to test the 4D EGB gravity.

\begingroup
\begin{table*}
	\caption{ 
		{\bf Estimation of Time Delay for SMBHs}: Estimation of  time delay between  the first and second relativistic image  for supermassive black holes at the center of nearby  galaxies when considered 4D EGB Bardeen black hole with $\tilde{\alpha}=0.5$ and $q=0.2$ in comparison with Schwarzschild and Bardeen black hole ($q=0.3$) . Mass ($M$) and distance ($D_{OL}$) are given in the units of solar mass and Mpc, respectively. Time delays are expressed in minutes. 
	}\label{Table2}
\resizebox{1\textwidth}{!}{
		\begin{tabular}{p{2cm} p{2cm} p{2cm} p{2cm} p{2cm} p{2cm} p{2cm}}
Galaxy & $M( M_{\odot})$ & $D_{OL}$ (Mpc) & $M/D_{OL}$ & $\Delta T^s_{12}$& $\Delta T^s_{12}$ & $\Delta T^s_{12}$\\
 &  &  &  & \text{Schwarzschild}& \text{Bardeen} & \text{4D EGB Bardeen} \\			
				\hline
Milky Way& $  4.0\times 10^6	 $ & $0.008 $ &       $2.471\times 10^{-11}$ & $10.69459 $ & $9.961004$ & $9.860882$     \\
M87&$ 6.5\times 10^{9} $&$ 16.8 $
&$1.758\times 10^{-11}$ & 17378.7 & 16186.6 & 16023.9 \\
NGC 4472 &$ 2.54\times 10^{9} $&$ 16.72 $
&$7.246\times 10^{-12}$& $6791.06$ & 6325.24 & 6261.66 \\
			 
NGC 1332 &$ 1.47\times 10^{9} $&$22.66  $
&$3.094\times 10^{-12}$& $3930.26$ & 3660.67 & 3623.87 \\
			
NGC 4374 &$ 9.25\times 10^{8} $&$ 18.51 $
&$2.383\times 10^{-12}$& $2473.12$ & 2303.48 & 2280.33 \\
			 
NGC 1399&$ 8.81\times 10^{8} $&$ 20.85 $
&$2.015\times 10^{-12}$& $2355.48$ & 2193.91 & 2171.86 \\
			 
NGC 3379 &$ 4.16\times 10^{8} $&$10.70$
&$1.854\times 10^{-12}$& $1112.24$ &  1035.94 & 1025.53 \\
			
NGC 4486B &$ 6\times 10^{8} $&$ 16.26 $
&$1.760\times 10^{-12}$ & $1604.19$ &  1494.15 & 1479.13 \\
			 
NGC 1374 &$ 5.90\times 10^{8} $&$ 19.57 $ &$1.438\times 10^{-12}$& $1577.45$ &  1469.25 & 1454.48 \\
			    
NGC 4649&$ 4.72\times 10^{9} $&$ 16.46 $
&$1.367\times 10^{-12}$& $12619$ & 11754. & 11635.8 \\
			
NGC 3608 &$  4.65\times 10^{8}  $&$ 22.75  $ &$9.750\times 10^{-13}$& $1243.25$ &  1157.97 & 1146.33 \\
			
NGC 3377 &$ 1.78\times 10^{8} $&$ 10.99$
&$7.726\times 10^{-13}$ & $475.909$ &  443.265 & 438.809 \\
			
NGC 4697 &$  2.02\times 10^{8}  $&$ 12.54  $ &$7.684\times 10^{-13}$& $540.077$ &   503.031 & 497.975 \\
			 
NGC 5128 &$  5.69\times 10^{7}  $& $3.62   $ &$7.498\times 10^{-13}$& $152.131$ &  141.695 & 140.271 \\
			 
NGC 1316&$  1.69\times 10^{8}  $&$20.95   $ &$3.848\times 10^{-13}$& $451.816 $ &  420.852 & 416.622 \\
			 
NGC 3607 &$ 1.37\times 10^{8} $&$ 22.65  $ &$2.885\times 10^{-13}$& $366.265 $ &  341.164 & 337.735 \\
			
NGC 4473 &$  0.90\times 10^{8}  $&$ 15.25  $ &$2.815\times 10^{-13}$& $240.628$ &  224.123 & 221.87 \\
			
NGC 4459 &$ 6.96\times 10^{7} $&$ 16.01  $ &$2.073\times 10^{-13}$ & $186.086 $ &   173.321 & 171.579 \\

M32 &$ 2.45\times 10^6$ &$ 0.8057 $
&$1.450\times 10^{-13}$ & $6.5504 $ &  6.10112 & 6.03979 \\
			 
NGC 4486A &$ 1.44\times 10^{7} $&$ 18.36  $ &$3.741\times 10^{-14}$ & $38.5005$ &  35.8596 & 35.4992 \\
			 
NGC 4382 &$  1.30\times 10^{7}  $&$ 17.88 $  &$3.468\times 10^{-14}$& $34.7574 $ &  32.3733 & 32.0479 \\
		\end{tabular}
}		
\end{table*}
\endgroup
Measuring the  lensing observables $\theta_{\infty}$, $s$ and $r_{\text{mag}}$ from the observation, one can find the coefficients $\bar{a}$ and $\bar{b}$ in SDL  \citep{Bozza:2002zj}. Comparing the calculated values with those predicted by the theoretical models, we can get  information about the parameters of the lens (black hole).

\section{Lensing by supermassive black holes} \label{sec6}
We model here the supermassive black holes Sgr A*  in our galactic center, M87* in Meisser 87 galaxy and several other  supermassive black holes as the 4D EGB Bardeen black holes for numerical estimation of lensing observables \citep{Virbhadra:2008ws,Poshteh:2018wqy,Khodabakhshi:2020ddv,Islam:2020xmy,Kumar:2020sag,Kumar:2021cyl}.  With the source distance $D_{OS}=2D_{OL}$, we estimate the angular position of the innermost image  $\theta_{\infty}$, the angular separation of the outermost image with the remaining bunch of relativistic images  $s$ and the relative magnification $r_{\text{mag}}$ in order to get some information about the charge $q$ and coupling constant $\tilde{\alpha}$ for 4D EGB Bardeen black holes. The behaviour of  $\theta_{\infty}$ and $s$ with the parameters $\tilde{\alpha}$ and $q$  for Sgr A* and M87* has been depicted in Fig.~\ref{plot8} and Fig.~\ref{plot9}, respectively.  These observables for  Sgr A* and M87*  are tabulated in Table \ref{table1} for the same set of observables when the lens is a Schwarzschild black hole, Bardeen black holes or 4D EGB black holes are also enlisted for comparison. Our results in limits $\tilde{\alpha}\to 0$ encompass those of Bardeen black holes  \citep{Eiroa:2010wm} and of Schwarzschild black holes \citep{Virbhadra:1999nm,Bozza:2002zj} when $\tilde{\alpha}\to 0,~q=0$ but if only $q=0$ we obtain the results of 4D EGB black holes \citep{Islam:2020xmy}. 
\begingroup
\begin{table*}
	\caption{ 
	{\bf Image positions of SMBHs}: Image positions  of first and second relativistic images  on the same side of the source for  supermassive black holes at the center of nearby  galaxies when considered 4D EGB Bardeen black hole with $\tilde{\alpha}=0.5$ and $q=0.2$ in comparison with Schwarzschild black hole and Bardeen black hole ($q=0.3$) at $\beta=1~$arrsec. }\label{Table3}
\resizebox{1\textwidth}{!}{
		\begin{tabular}{p{2cm} p{2cm} p{2cm} p{2cm} p{2cm} p{2cm} p{2cm}}
			\multicolumn{1}{c}{Galaxy}&
			\multicolumn{2}{c}{Schwarzschild Black hole}&
			\multicolumn{2}{c}{Bardeen Black hole}&
			\multicolumn{2}{c}{4D EGB Bardeen Black hole}\\
		    
			%%%%%%%%%%%%%%%%%%%%%%%%%%%%%%%%%%%%% 
&$\theta_{1p}$  & $\theta_{2p}$   & $\theta_{1p}$  & $\theta_{2p}$ & $\theta_{1p}$  & $\theta_{2p}$ \\
           \hline
			%%%%%%%%%%%%%%%%%%%%%%%%%%%%%%%%%%%%%
Milky Way & 25.59627 & 25.56433 & 23.87557 & 23.81099 & 23.67115 & 23.57215 \\
			
M87 & 19.8066 & 19.7819 & 18.4751 & 18.4252 & 18.317 & 18.2404 \\	
 
NGC 4472 & 7.77686 & 7.76715 & 7.25406 & 7.23444 & 7.19195 & 7.16187 \\
			
NGC 1332 & 3.32096 & 3.31682 & 3.09771 & 3.08933 & 3.07119 & 3.05835 \\
			
NGC 4374  &  2.55824 & 2.55505 & 2.38627 & 2.37981 & 2.36584 & 2.35594 \\
			
NGC 1399  & 2.1631 & 2.1604 & 2.01769 & 2.01223 & 2.00041 & 1.99204 \\
			
NGC 3379 &  1.99029 & 1.98781 & 1.85649 & 1.85147 & 1.8406 & 1.8329 \\
			
NGC 4486B &  1.88902 & 1.88667 & 1.76203 & 1.75727 & 1.74695 & 1.73964 \\
			
NGC 1374  &  1.54336 & 1.54144 & 1.43961 & 1.43572 & 1.42728 & 1.42132 \\
			
NGC 4649  & 14.6798 & 14.6614 & 13.6929 & 13.6559 & 13.5757 & 13.5189 \\
			
NGC 3608 & 1.04635 & 1.04505 & 0.976012 & 0.973372 & 0.967656 & 0.963609 \\
			
NGC 3377 & 0.829142 & 0.828108 & 0.773403 & 0.771311  & 0.766781 & 0.763575 \\

NGC 4697 & 0.824633 & 0.823604 & 0.769197 & 0.767116  & 0.762611 & 0.759422 \\

NGC 5128 & 0.804656 & 0.803652 & 0.750564 & 0.748533 & 0.744137 & 0.741025 \\
			 
NGC 1316& 0.412961 & 0.412446 & 0.3852 & 0.384158 & 0.381902 & 0.380305 \\
			 
NGC 3607 &0.309641 & 0.309255 & 0.288826 & 0.288045 & 0.286353 & 0.285155 \\
			
NGC 4473 &0.30212 & 0.301743 & 0.28181 & 0.281048 & 0.279397 & 0.278229 \\
			
NGC 4459 &0.222548 & 0.222271 & 0.207588 & 0.207026 & 0.20581 & 0.20495 \\
			
M32 & 0.155668 & 0.155474 & 0.145203 & 0.144811 & 0.14396 & 0.143358 \\
			 
NGC 4486A  &0.040151 & 0.0401009 & 0.0374519 & 0.0373506 & 0.0371312 & 0.0369759 \\
			 
NGC 4382 & 0.0372205 & 0.0371741 & 0.0347184 & 0.0346245 & 0.0344211 & 0.0342772 \\

		\end{tabular}
}		
\end{table*}
\endgroup
\begingroup
\begin{table*}
	\caption{ 
	{\bf Magnification of SMBHs}:  Magnifications of first and second relativistic images  on the same side of the source for  Supermassive black holes at the center of nearby  galaxies when considered 4D EGB Bardeen black hole with $\tilde{\alpha}=0.5$ and $q=0.2$ in comparison with Schwarzschild and Bardeen black hole ($q=0.3$)
	at $\beta=1~$arrsec. }\label{Table4}
\resizebox{1\textwidth}{!}{
		\begin{tabular}{p{2cm} p{2cm} p{2cm} p{2cm} p{2cm} p{2cm} p{2cm}}
		
			\multicolumn{1}{c}{Galaxy}&
			\multicolumn{2}{c}{Schwarzschild Black hole}&
			\multicolumn{2}{c}{Bardeen Black hole}&
			\multicolumn{2}{c}{4D EGB Bardeen Black hole}\\
		    
			%%%%%%%%%%%%%%%%%%%%%%%%%%%%%%%%%%%%% 
     	& $\mu_{1p}$ &    $\mu_{2p}$  &  $\mu_{1p}$ &    $\mu_{2p}$ \\
			\hline
			%%%%%%%%%%%%%%%%%%%%%%%%%%%%%%%%%%%%%
Milky Way & $ 7.94126\times 10^{-18}$ & $1.48134 \times 10^{-20}$ &   $1.30512\times10^{-17} $&$ 5.528328\times10^{-20}$& 
$1.772956\times 10^{-17}$ &  $1.361426\times 10^{-19}$ \\

M87 &$4.75508\times 10^{-18}$ & $8.869999\times 10^{-21}$ &
$7.814811\times10^{-18} $& $3.310259\times10^{-20}$&
$1.061613\times 10^{-17} $&$8.151964\times 10^{-20}$\\	

NGC 4472 & $7.33069\times 10^{-19}	 $& $  1.36745\times 10^{-21}$ &  $1.204773\times10^{-18} $& $5.103272\times10^{-21}$&
$1.63664\times 10^{-18}$ &  $1.25675\times 10^{-20}$ \\ 

NGC 1332 &  $1.33679\times 10^{-19}	 $& $  2.49362\times 10^{-22}$ & $2.196975\times10^{-19} $& $9.306121\times10^{-22}$&
$2.98451\times 10^{-19}$ &  $2.29176\times 10^{-21}$ \\

NGC 4374  & $7.93269\times 10^{-20}	 $& $  1.47974\times 10^{-22}$ &   $1.30371\times10^{-19}  $& $5.522357\times10^{-22}$&
$1.77104\times 10^{-19}$ &  $1.35996\times 10^{-21}$  \\

NGC 1399  & $5.67139\times 10^{-20}	 $& $  1.05793\times 10^{-22}$ &  $9.320732\times10^{-20} $& $3.948149\times10^{-22}$&
$1.26619\times 10^{-19}$ &  $9.72286\times 10^{-22}$ \\

NGC 3379 & $4.80141\times 10^{-20}	 $& $ 8.95642\times 10^{-23}$ &  $7.890953\times10^{-20} $& $3.342512\times10^{-22}$&
$1.07196\times 10^{-19}$ &  $ 8.23139\times 10^{-22}$\\

NGC 4486B & $4.32525\times 10^{-20}	 $& $ 8.0682\times 10^{-23}$ &   $7.108398\times10^{-20} $& $3.011031\times10^{-22}$&
$9.65649\times 10^{-20}$ &  $7.41507\times 10^{-22}$\\

NGC 1374  & $2.88717\times 10^{-20}	 $& $ 5.38565\times 10^{-23} $ &  $4.744962\times10^{-20} $& $2.009908\times10^{-22}$&
$6.44585\times 10^{-20}$ &  $4.94967\times 10^{-22}$\\

NGC 4649  & $2.61201\times 10^{-18}	 $& $ 4.87236\times 10^{-21}$ &    $4.292741\times10^{-18} $& $1.818353\times10^{-20}$&
$5.83153\times 10^{-18}$ &  $ 4.47794\times 10^{-20}$\\

NGC 3608  & $1.327067\times 10^{-20}$ & $2.475475\times 10^{-23}$ &   $2.180989\times10^{-20} $& $9.238404\times10^{-23}$&
$2.962791\times 10^{-20}$ &  $ 2.275083\times 10^{-22}$\\

NGC 3377  & $8.332865\times 10^{-21}$ & $1.55439\times 10^{-23}$ &   $1.369478\times10^{-20} $& $5.800941\times10^{-23}$&
$ 1.860384\times 10^{-20}$ &  $ 1.428561\times 10^{-22}$\\

NGC 4697 & $8.24247\times 10^{-21}$ & $1.537528\times 10^{-23}$ &   $1.354622\times10^{-20} $& $5.738013\times10^{-23}$&
$1.840202\times 10^{-20}$ &  $ 1.413064\times 10^{-22}$\\

NGC 5128 &$7.847966\times 10^{-21}$ & $1.463938\times 10^{-23}$&  
 $1.289786\times10^{-20} $& $5.463378\times10^{-23}$&
 $1.752126\times 10^{-20}$ &  $ 1.345431\times 10^{-22}$\\
 
NGC 1316 &$2.067069\times 10^{-21}$ & $3.855856\times 10^{-24}$ &    $3.397158\times10^{-21} $& $1.438995\times10^{-23}$&
$4.61491\times 10^{-21}$ &  $ 3.543721\times 10^{-23}$\\

NGC 3607 & $1.162129\times 10^{-21}	$ & $2.167805\times 10^{-24}$ &   $1.909919\times10^{-21} $& $8.090187\times10^{-24}$&
$ 2.594553\times 10^{-21}$ &  $ 1.992319\times 10^{-23}$\\

NGC 4473 & $1.106356\times 10^{-21}	 $& $ 2.063767\times 10^{-24}$ &    $1.818259\times10^{-21} $& $7.701923\times10^{-24}$&
$2.470036\times 10^{-21}$ &  $ 1.896704\times 10^{-23}$\\

NGC 4459 & $6.003238\times 10^{-22}	 $& $ 1.119828\times 10^{-24}$ &    $9.866114\times10^{-22} $& $4.179166\times10^{-24}$&
$1.340274\times 10^{-21}$ &  $ 1.029177\times 10^{-23}$ \\

M32 & $2.937211\times 10^{-22}	 $& $ 5.478995\times 10^{-25}$ &   
$4.827205\times10^{-22} $& $2.044746\times10^{-24}$&
$6.557576\times 10^{-22}$ &  $  5.035465\times 10^{-24}$\\

NGC 4486A  & $1.954022\times 10^{-23}$ & $3.644979\times 10^{-26}$ &  $3.211367\times10^{-23}$ & $1.360296\times10^{-25}$&
$3.748949\times 10^{-23}$ &  $ 2.878762\times 10^{-25}$\\

NGC 4382 & $1.679196\times 10^{-23}	 $& $ 3.132327\times 10^{-26}$&  $2.759701\times10^{-23}$ & $1.168976\times10^{-25}$&
$3.748949\times 10^{-23}$ &  $ 2.878762\times 10^{-25}$    \\
		\end{tabular}
}		
\end{table*}
\endgroup
It is evident from Table \ref{table1} that for small values of $\tilde{\alpha}$ and $q$, the observational predictions of 4D EGB Bardeen black holes are indistinguishable from the Schwarzschild black hole, Bardeen black holes or 4D EGB black holes. However, the deviation becomes significant for large $\tilde{\alpha}$ and $q$,  as  Table \ref{table1} indicates that $\theta_{\infty}$ for 4D EGB Bardeen black holes  is always smaller and hence can be potentially differentiated from the GR counterparts provided we have a telescope with very strong resolving power. As an example, the deviation of $\theta_{\infty}$ for Sgr A* if considered a 4D EGB Bardeen black hole ($\tilde{\alpha}=0.5,~q=0.2$) from Bardeen black hole ($q=0.3$) is $0.2393~\mu$as whereas for M87*, the deviation is  $0.18519~\mu$as. Similarly, the deviation from 4D EGB black hole ($\tilde{\alpha}=0.95$) can reach up to $0.3074~\mu$as  for Sgr A* and  $0.237874~\mu$as for M87*.  In Table~\ref{Table3}, we give an estimation of  $\theta_{1p}$  and $\theta_{2p}$, which are the  angular positions of  first and second order primary images \citep{Virbhadra:2008ws} for black holes in several nearby galaxies which are arranged according to  decreasing ratio of $M/D_{OL}$,  by considering these black holes as Schwarzschild black hole, Bardeen black hole  ($q=0.3$) and  4D EGB Bardeen black hole with $\tilde{\alpha}=0.5$ and $q=0.2$. The deviation from GR are of the $\mathcal{O}(\mu)$as and such a deviation in a realistic astrophysical environment is certainly not feasible in the near future, however if the images can be resolved, it would provide an excellent test of gravity in  SDL. Also, as suggested from Fig.~\ref{plot11},  $\theta_{1p} > |\theta_{1s}|$ for higher values of $\beta$ \citep{Poshteh:2018wqy,Khodabakhshi:2020ddv}.  Further, the separation $s$ due to 4D EGB Bardeen black holes for Sgr A* and M87* ranges between 0.03199-0.148959 $\mu$as and 0.0247-0.11526 $\mu$as, respectively. If  these images could be resolved, it is possible to calculate their magnification. The absolute  magnification  of the first and second order images are estimated in Table~\ref{Table4} for black holes in nearby galaxies by considering them as  4D EGB Bardeen black hole with $\tilde{\alpha}=0.5$ and $q=0.2$ and compared with the Schwarzschild black hole \citep{Virbhadra:1999nm,Virbhadra:2008ws} and Bardeen black hole  ($q=0.3$) \citep{Eiroa:2010wm}. The first order images of 4D EGB Bardeen black holes are highly magnified than the second order images as well as the corresponding images in GR. The ratio of the flux from the first image to all other images  $\in$  (4.65751,\; 6.82173), however, decreases with  $\tilde{\alpha}$ and $q$ (cf. Fig.~\ref{plot10}).

\begin{figure*}
	\begin{centering}
		\begin{tabular}{c c}
	     	\includegraphics[scale=0.75]{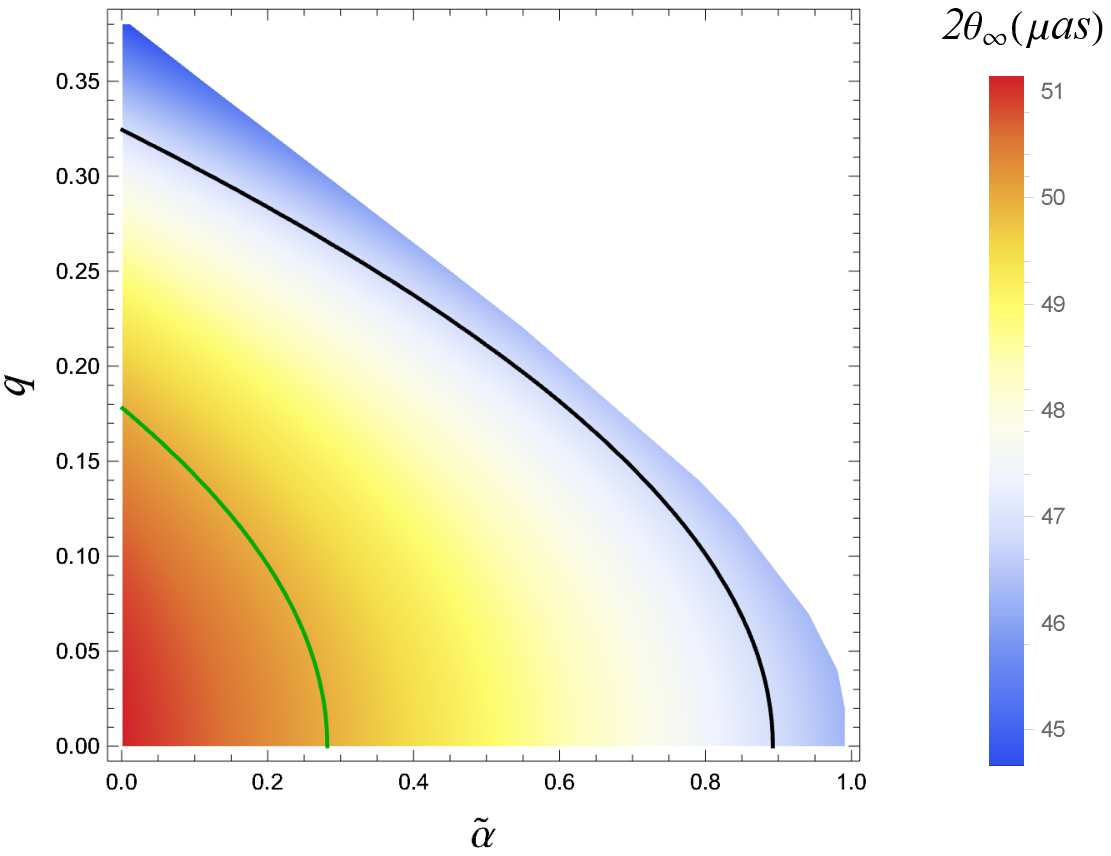}
		    \includegraphics[scale=0.75]{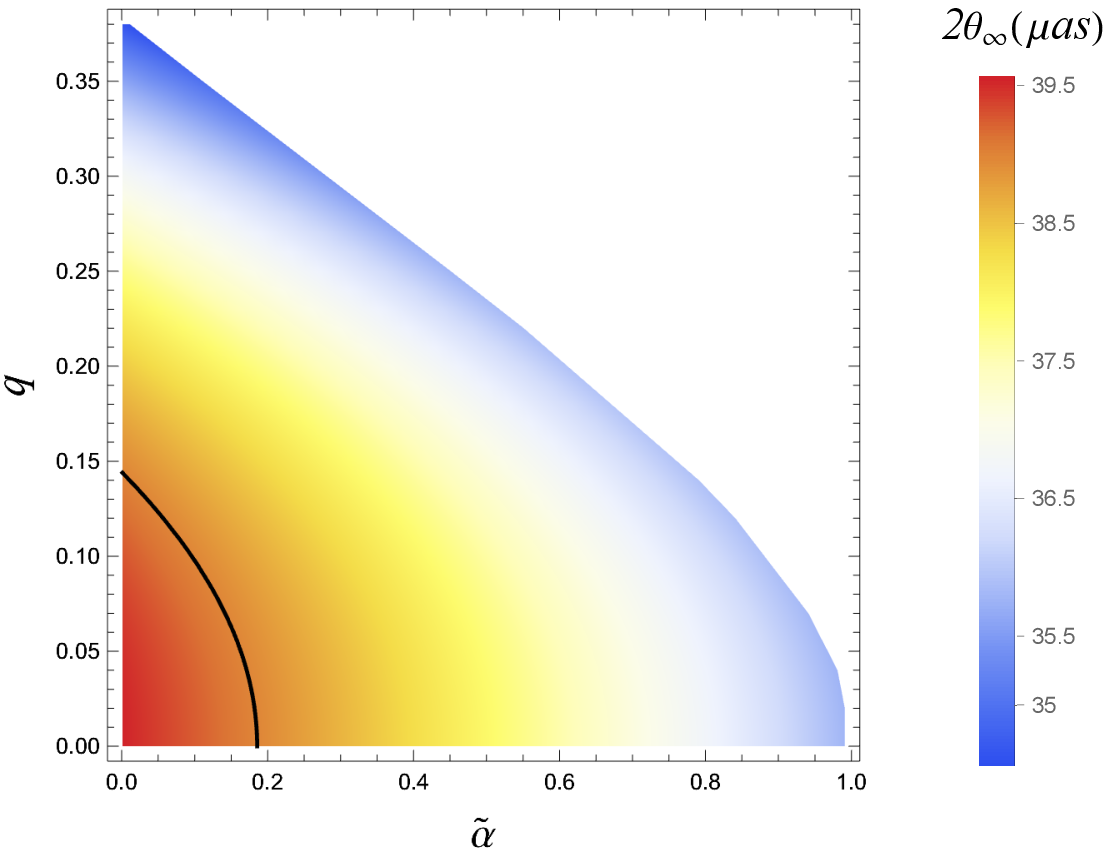}&\\
			\end{tabular}
	\end{centering}
	\caption{Shadow angular diameter $\theta_{sh}=2\theta_{\infty}$ of 4D EGB Bardeen black holes as a function of $(\tilde{\alpha},q)$. The black and green lines correspond to the Sgr A* black hole shadow at $\theta_{sh}=46.9~\mu$as and $\theta_{sh}=50~\mu$as, respectively, such that the region between these lines satisfies the Sgr A* shadow 1-$\sigma$ bound (\textit{left}). M87* shadow angular diameter when considered as a 4D EGB Bardeen black hole. The black line is $\theta_{sh}=39~\mu$as, and the region contained within it meets the M87* shadow 1-$\sigma$ bound (\textit{right}). } \label{plot13}		
\end{figure*} 
\section{Constraints from EHT observations of M87* and Sgr A*}
We use the EHT observation results of M87* and Sgr A* black hole shadows to constrain the black hole parameters associated with 4D EGB Bardeen black holes. 
By considering the apparent radius of the photon sphere ($\theta_{\infty}$) as the angular size of the black hole shadow, we  constraint the parameters ($\tilde{\alpha}$, $q $) within the 1-$\sigma$ level.   We model the M87* and Sgr A* as the Bardeen black holes and use the M87* and Sgr A* shadow results to test the viability of these models to explain the astrophysical black hole spacetimes. 
\subsubsection{Constraints from  M87* }
The EHT collaboration delivered the first image of the supermassive black hole M87* in 2019, producing a ring of diameter $\theta_d= 42\pm 3\,\mu$as \citep{Akiyama:2019cqa1}. We find that 
the Schwarzschild black hole  ($\tilde{\alpha}=0$, $q = 0$) casts the largest shadow with its angular diameter $\theta_{sh}=2\theta_\infty= 39.6192~\mu$as, which falls within the 1-$\sigma$ region for the black hole with mass  $M=(6.5 \pm 0.7) \times 10^9 M_\odot$ and distance of $D_{OL}=16.8$ Mpc \citep{EventHorizonTelescope:2019dse,EventHorizonTelescope:2019pgp,EventHorizonTelescope:2019ggy}. Fig.~\ref{plot13} depicts the angular diameter $\theta_{sh}$ as a function of ($\tilde{\alpha},q $), with the black solid line corresponding to $\theta_{sh}=39~\mu$as for the 4D EGB Bardeen black holes as M87*. The 4D EGB Bardeen black hole metric when investigated with the EHT results of M87*  within the 1-$\sigma$ bound, constrains the parameters ($\tilde{\alpha}, q$), viz., $0< \tilde{\alpha}\le 0.185644  $ and $0< q \le 0.14427 $. Thus, based on Fig.~\ref{plot13}, 4D EGB Bardeen black holes can be a candidate for the astrophysical black holes.

\subsubsection{Constraints from  Sgr A* shadow}
In contrast to the EHT results of M87* black hole, the EHT result for Sgr A* not only calculated the emission ring angular diameter $\theta_d=(51.8\pm 2.3)\mu$as but also estimated the shadow diameter $\theta_{sh}=(48.7\pm 7)\mu$as with the prior perceived estimates  $M=4.0^{+1.1}_{-0.6} \times 10^6 M_{o}$ and $D_{LS}=8.15\pm 0.15$ kpc \citep{EventHorizonTelescope:2022xqj}.   The EHT observation used three independent algorithms, eht-imaging, SIMLI, and DIFMAP, to find out that the averaged measured value of the shadow angular diameter lies within range $\theta_{sh} \in (46.9, 50)~\mu$as, and the 1-$\sigma$ interval is $\in$ $(41.7 ,55.6)~\mu$as. The angular diameter  $\theta_{sh} \in (46.9, 50) \mu$as, which falls within the $1- \sigma$ confidence region with the observed angular diameter of the EHT observation of Sgr A* black hole strongly constrains the parameters $0.178241 \le q \le 0.324535$ and $0.281766 \le \tilde{\alpha} \le 0.892415$  for the 4D EGB Bardeen black hole.  Thus, within the finite parameter space, 4D EGB Bardeen black holes definitively agree with the EHT results of Sgr A* black hole shadow (cf. Fig.~\ref{plot13}).

\section{conclusion}\label{sect6}
We analyzed strong gravitational lensing by 4D EGB black holes to assess the dependence of observables, deflection angle and time delay on the parameters $\tilde{\alpha}$ and $q$. Beginning with the Euler-Lagrangian formalism, we showed that the particle's trajectory is indeed influenced by the coupling constant $\tilde{\alpha}$ and NED charge $q$. Interestingly, the 4D EGB Bardeen black holes make smaller photon spheres compared to  Bardeen black holes or 4D EGB black holes, and the unstable photon orbit radius is a decreasing function of $\tilde{\alpha}$ and $q$. Moreover, the dependence of critical impact parameters  $u_m$ on $\tilde{\alpha}$  and  $q$  has decreasing behaviours which are qualitatively the same as that of the unstable photon orbit radius $x_m$ and its value is always smaller than the GR counterparts. The deflection angle $\alpha_D(\theta)$, for fixed impact parameter $u$, is always higher for the Schwarzschild black hole. Further, the photon makes its first loop around the 4D EGB Bardeen black hole with $\tilde{\alpha}=0.5$ and $q=0.2$ at $u=2.40566$   whereas the corresponding value for Schwarzschild black hole and Bardeen black hole ($q=0.3$) is  $u=2.601309$ and $2.426437$, respectively. The lensing coefficients $\bar{a}$ increases, while  $\bar{b}$ decreases with $\tilde{\alpha}$ and  $q$.

The angular position $\theta_{n}$ of the images in 4D EGB Bardeen black holes is more diminutive than Schwarzschild and Bardeen black holes. The  $\theta_{\infty} $ rapidly decreases,  but the angular separation between the first and innermost image  $s$ is higher when compared to the Schwarzschild black hole or Bardeen black holes. The $\theta_{\infty}$ ranges between 23.1853-25.56427  $\mu$as for Sgr A* and its maximum deviation from GR counterpart can reach up to  2.3789 $\mu$as. For M87*, it ranges between  17.941 - 19.7819 $\mu$as and deviation is as much as  1.84084$\mu$as. The separation $s$, an increasing function of $\tilde{\alpha}$ and $q$, due to 4D EGB Bardeen black holes for Sgr A* and M87* range between 0.031997-0.14895 $\mu$as and 0.0247-0.1152 $\mu$as, respectively. Although, the ratio of the flux from the first image to all other images, decreases with  $\tilde{\alpha}$ and $q$ and $\in$ (4.65751, 6.82173),  the first order images of 4D EGB Bardeen black holes are highly magnified than the second order images as well as the corresponding images in GR.  Furthermore, the time delay between the first and second-order images for Sgr A* and M87* black holes when considered  4D EGB Bardeen black holes, respectively,  can reach $\sim9.86088$~min and $\sim16023.93$~min such that the time delay difference between 4D EGB Bardeen black holes and Schwarzschild black holes, respectively,  is $0.8337$~min and $1354.775$~min.
These differences in time delays, except for Sgr A*, is significant enough for astronomical measurements, provided we have enough angular resolution separating two relativistic images.

The results presented here generalise previous discussions on  black holes lensing in GR viz. Schwarzschild,  Bardeen and 4D EGB gravity black holes contained, respectively, in the limits, $\alpha, q \to 0$,  $\alpha \to 0 $, and $q \to 0$.  Although it is tough to resolve the order estimated in SDL, the outlook of future observations looks bright. The Event Horizon Telescope observation of M87* has achieved angular resolution of 20 $\mu$as. Thus, it is essential to use GR and other alternate theories of gravity to give a realistic view of the observed images.  Thus, we have considered the first regular black hole metric, viz the Bardeen metric, to deviate from its Kerr metric to show that we can get significant constraints with the 2017 EHT results of M87* and Sgr A*. 

Finally, due to the complicated higher-order curvature metric ~(\ref{fr}), in the present analysis, we have restricted our study to the spherically symmetric case, i.e.,  overlooked  spin. This is because, although the radii of the photon orbits strongly depend on the spin, the silhouette of the shadow, as observed at infinity, has a size and a shape weakly depending on the spin of the black hole \citep{Bardeen:1973tla,EventHorizonTelescope:2020qrl}.  However,  we can reasonably expect our results on lensing by supermassive black holes Sgr A* and M87* are valid,  and the EHT observation can also be adopted to test these spherical black holes. Meanwhile, some detailed investigation for the rotating counterpart will be a promising avenue for the future.
\section{Acknowledgments} 
S.U.I and S.G.G. would like to  thank SERB-DST for the project No. CRG/2021/005771. S.D.M acknowledges that this work is based upon research supported by the South African Research Chair Initiative of the Department of Science and Technology and the National Research Foundation.
\section{DATA AVAILABILITY}
This is entirely theoretical work, and all of the results presented in the manuscript are derived from the equations. We did not generate any original data during the course of this study, nor did we analyse any third-party data in this article.
\bibliographystyle{mnras}
\bibliography{Bardeen1}
%%%%%%%%%%%%%%%%%%%%%%%%%%%%%%%%%%%%%%%%%%%%%%%%%%
%%%%%%%%%%%%%%%%%%%%%%%%%%%%%%%%%%%%%%%%%%%%%%%%%%
% Don't change these lines
\bsp	% typesetting comment
\label{lastpage}
\end{document}